\documentclass[english,twocolumn,transaction]{IEEEtran}
\usepackage[T1]{fontenc}
\usepackage{float}
\usepackage{amsmath}
\usepackage{amsthm}
\usepackage{amssymb}
\usepackage{stackrel}
\usepackage{graphicx}
\usepackage{setspace}
\usepackage{url}
\usepackage{color}
\usepackage{lipsum} 

\usepackage{caption}
\usepackage{subfigure}
\usepackage{algorithm}
\usepackage{algorithmic}

\makeatletter

\floatstyle{ruled}
\newfloat{algorithm}{tbp}{loa}
\providecommand{\algorithmname}{Algorithm}
\floatname{algorithm}{\protect\algorithmname}

\theoremstyle{plain}

\theoremstyle{definition}

\theoremstyle{plain}

\theoremstyle{plain}

\IEEEoverridecommandlockouts
\usepackage{ifpdf}
\usepackage{cite}
\ifCLASSOPTIONcompsoc
\usepackage[caption=false,font=normalsize,labelfont=sf,textfont=sf]{subfig}
\else
\usepackage[caption=false,font=footnotesize]{subfig}
\fi
\usepackage{amsmath}
\usepackage{mathtools}
\usepackage{booktabs}
\usepackage{multirow}
\usepackage{setspace}

\@ifundefined{showcaptionsetup}{}{
 \PassOptionsToPackage{caption=false}{subfig}}
\usepackage{subfig}
\makeatother

\usepackage{babel}

\renewcommand\figurename{Fig.}

\begin{document}
\captionsetup[figure]{font={small}, name={Fig.}, labelsep=period}

\title{Complex ResNet Aided DoA Estimation for Near-Field MIMO Systems}
\author{Yashuai Cao,~\IEEEmembership{Student Member,~IEEE}, Tiejun Lv,~\emph{Senior Member, IEEE},\\
Zhipeng Lin, Pingmu Huang, and~Fuhong Lin

\thanks{The financial support of the National Natural Science Foundation of China
(NSFC) (Grant No. 61671072 and 61931001) and the Beijing Natural Science Foundation (No. L192025) is gratefully acknowledged.
(\emph{Corresponding author: Tiejun Lv.})

Y. Cao, T. Lv, Z. Lin and P. Huang are with the School of Information and Communication Engineering, Beijing University of Posts and Telecommunications (BUPT), Beijing 100876, China (e-mail: \{yashcao, lvtiejun, linlzp, pmhuang\}@bupt.edu.cn).

F. Lin is with the School of Computer and Communication Engineering, University of Science and Technology Beijing,
Beijing 100083, China (e-mail: FHLin@ustb.edu.cn).
}}

\maketitle
\begin{abstract}
The near-field effect of short-range multiple-input multiple-output (MIMO) systems imposes many challenges on direction-of-arrival (DoA) estimation.
Most conventional scenarios assume that the far-field planar wavefronts hold. In this paper, we investigate the DoA estimation problem in short-range MIMO communications, where the effect of near-field spherical wave is non-negligible.
By converting it into a regression task, a novel DoA estimation framework based on complex-valued deep learning (CVDL) is proposed for the near-field region in short-range MIMO communication systems.
Under the assumption of a spherical wave model, the array steering vector is determined by both the distance and the direction.
However, solving this regression task containing a massive number of variables is challenging, since datasets need to capture numerous complicated feature representations.
To overcome this, a virtual covariance matrix (VCM) based on received signals is constructed, and thus such features extracted from the VCM can deal with the complicated coupling relationship between the direction and the distance.
Although the emergence of wireless big data driven by future communication networks promotes deep learning-based wireless signal processing, the learning algorithms of complex-valued signals are still ongoing.
This paper proposes a one-dimensional (1-D) residual network that can directly tackle complex-valued features due to the inherent 1-D structure of signal subspace vectors.
In addition, we put forth a cropped VCM based policy which can be applied to different antenna sizes.
The proposed method is able to fully exploit the complex-valued information. Our simulation results demonstrate the superiority of the proposed CVDL approach over the baseline schemes in terms of the accuracy of DoA estimation.
\end{abstract}

\renewcommand\figurename{Fig.}
\newcommand{\tabincell}[2]{\begin{tabular}{@{}#1@{}}#2\end{tabular}}

\section{Introduction}
With the advances of key enabling technologies in the fifth generation (5G) mobile communications, such as millimeter wave (mmWave), small cell, and massive multiple-input multiple-output (MIMO), the direction-of-arrival (DoA) estimation problem \cite{pillai2012array} in the context of massive MIMO systems has attracted substantial research in recent years \cite{koivisto2017high}.
In past decades, many DoA estimation methods have been proposed, varying from subspace-based techniques \cite{6777542, LV201630, 7933068}, maximum-likelihood estimators \cite{miller1990maximum}, compressed sensing methods\cite{yu2009compressive}, to hybrid methods \cite{carlin2013directions, liu2012efficient, el1997performance}.
Efficient and accurate DoA estimation algorithms are crucial to massive MIMO systems \cite{7523373, 8377155, 8254891}.
Specifically, accurate DoA information is shown to be critical for beamforming design in massive MIMO systems \cite{wang2015two}.

Most of DoA estimation methods mentioned above assume far-field planar wavefront. However, this assumption may be invalid for future mobile systems.
As a promising enabling technique for 5G-and-beyond mobile networks, mmWave massive MIMO is expected to support high data-rate communications.
However, the short wavelength and large antenna aperture induce challenges in the modeling of propagation patterns.
When the Rayleigh distance is considered, the last-order scatterers fall within the near-field region of the receive antenna, where spherical wave propagation is applicable \cite{yin2017scatterer}.
On the other hand, the near-field DoA estimation appears to be vitally important in many other applications, including vehicle-to-everything (V2X) communications \cite{8108347, 6692160}, virtual reality, smart home, and automated driving.
Particularly, the DoA estimation of signals from landmarks is the core technology of vehicle positioning \cite{9031722}, which is an important branch in the Internet of Vehicles (IoV).

\subsection{Related Work and Motivation}

In near-field DoA estimation, the joint estimation of the distance and the angle is inevitable, which extends the search space and increases the search overhead. To overcome this issue, a subspace method without multidimensional search is proposed in \cite{6509484}. These subspace methods require many snapshots and high signal-to-noise ratio (SNR) to improve the spatial resolution. Furthermore, a maximum-likelihood-based approach \cite{huang1991near} by iterative estimation is proposed for the case of a small number of snapshots and low SNR. However, as the parameter dimension increases, the iterative methods inevitably face the predicament of slow processing speed.

Meanwhile, wireless networks are generating a massive volume of data, from which we can extract useful information using the state-of-the-art machine learning (ML) techniques \cite{li2017wireless, mlwc8360430, 8382166, li2019survey}.
There have been some efforts exploring the applications of ML in DoA estimation.
In \cite{donelli2009innovative}, the near-field DoA estimation is modeled as a classification problem based on the support vector machine (SVM) approach.
The authors of \cite{liu2018direction} extend the application of deep neural networks (DNNs) to a general far-field acoustic DoA estimation problem, which is modeled as a classification problem and not suitable for near-field estimation.
Although these ML-based classification methods are interesting, their accuracy depends on the quantization resolution of angles.
In \cite{7848281}, a support vector regression (SVR) method is proposed for estimating the DoAs of near field sources.
Popular deep learning (DL) methods are also used in far-field DoA estimation such as in the field of acoustic source direction finding \cite{chakrabarty2018multi}. However, such models of acoustic DoA estimation cannot be directly applied to signal processing in wireless systems.
In addition, a real-valued DNN (RVNN) for estimating DoA in hybrid massive MIMO channels is designed in \cite{8845653}. Traditionally, complex-valued signals are split into the real and imaginary parts, which are real values and fed into the network \cite{huang2018regression, liu2018direction, 8845653, 8903003}. Nonetheless, this treatment may fail to capture the correlation between the real part and the imaginary part, thus incurring the phase information loss.

Most of the existing DL methods focus on the real domain. However, many wireless communications problems to be solved are in the complex domain. Merging the conception of complex-valued neural network (CVNN) with wireless signal processing may provide a further thrust to enable smart radio.
The complex-valued deep learning (CVDL) methods refer to the DNNs that can perform complex arithmetic, and the weight parameters of CVNNs are complex-valued \cite{127967}, \cite{kim2002fully}.
The authors of \cite{zhang2017complex,trabelsi2017deep,scardapane2018complex} explored the applications of CVDL in image and speech processing.
These network architectures are all designed to account for image classification, music transcription or speech spectrum prediction.
Against the above backdrop, our motivations can be explained from the following aspects:
\begin{itemize}
\item Since the conventional near-field methods require high SNR and snapshots or online iterative estimation, it is unlikely to estimate DoA in real time. Moreover, there is a very limited study about designing DL methods for near-field DoA estimation. As DL is a powerful tool to solve intractable nonlinear problems, it would be interesting to improve the performance robustness if training an end-to-end network to solve the near-field DoA estimation.
\item Both spatial spectrum search and classification-based ML methods cannot meet the precision requirements when using coarse grids. Besides, RVNNs may destroy the structure of complex data via splitting real and imaginary parts. Thus, a CVNN-based regression model is expected to address these issues.
\item When near-field signals are concerned, the CVDL-based DoA estimation is not straightforward due to the coupling between the DoA and range parameters. How to design input features, activation functions and loss metrics motivates us to pursue a general CVDL-based DoA estimation framework.
\end{itemize}

\subsection{Main Contributions}

In this paper, a novel CVDL-aided DoA estimation approach is proposed for near-field MIMO systems, where the DoA estimation is viewed as an end-to-end regression task rather than a classification problem. First, to decouple the distance and the direction, we design a new feature representation based on the reconstructed virtual covariance matrix (VCM). This representation can eliminate the learning obstacles in DoA estimation.
Then, a complex-valued residual network (ResNet) is proposed to learn complex features of the input data. Compared with the RVNN, the proposed network circumvents the trivial split and splice between real and imaginary parts.
To the best of our knowledge, it is the first time that the phase mapping activation function is developed for the CVNN. How to introduce CVDL into massive MIMO systems has not been well addressed before.


The main contributions of this paper are summarized as follows:
\begin{itemize}
\item From an ML perspective, we investigate the near-field DoA estimation problem, which is considered as a phase feature extraction process of input data. Without the need for tedious grid-based classification, we train an end-to-end deep network for angle estimation. This network is capable of processing complex-valued tasks directly.
\item New features, based on the reconstructed VCM instead of the whole or half covariance matrix, are designed as the inputs of the CVNN. Therefore, the computational complexity of our deep networks is greatly decreased. Moreover, the joint two-dimensional (2-D) parameter estimation problem in the near-field is transformed into a one-dimensional (1-D) DoA estimation problem, where the designed input features are able to eliminate the complicated coupling relationships between the distance and the direction.
\item A 1-D convolutional complex-valued ResNet is proposed to process 1-D complex features. We design a residual block architecture with a shortcut connection to ensure the training stability. To map the real-valued angles from the complex-valued output, we devise a phase mapping activation function specifically for the DoA estimation task. Besides, we also carefully select different complex-valued activation functions in the corresponding neuron layers according to the distribution of the dataset.
\item For our specific regression task, we compare the training results of different loss metrics and determine the suitable loss function through comparison. As a result, the prediction precision is enhanced.
\item To realize a general CVNN model which can be applicable to different antenna sizes, we put forth a cropped VCM based scheme. This scheme can be deployed for arbitrary antenna size without the need for redesigning new networks. Moreover, it can reduce the computational cost since the inputs are cropped to a fixed shape.
\end{itemize}

The remainder of this paper is organized as follows. In Section \ref{section:system}, we establish the received signal model for the near-field region. In Section \ref{section:train}, the VCM reconstruction method is proposed to remove the corruption of range parameters on the signal subspace; and the feature preprocessing scheme is proposed based on the VCM. In Section \ref{section:cvnn}, we present the CVDL-aided DoA estimation framework and introduce the complex ResNet with specific model settings. Our simulation results are provided in Section \ref{section:sim}, which demonstrates that the CVNNs outperform the baseline methods. Our conclusions are offered in Section \ref{section:con}.

\emph{Notations}: Lower-case boldface symbols indicate column vectors, and upper-case boldface symbols refer to matrices; $(\cdot)^{\ast}$, $(\cdot)^{\mathrm{T}}$, and $(\cdot)^{\mathrm{H}}$ represent the conjugate, the transpose, and the conjugate transpose; $[\cdot]_{p,q}$ represents the entry on the $p$-th row and $q$-th column of a matrix; $j=\sqrt{-1}$ stands for the imaginary unit; $\lfloor\cdot\rfloor$ denotes the floor function; $\mathbb{E}(\cdot)$ represents the expectation of random variables; $\Re(\cdot)$ and $\Im(\cdot)$ indicate the real and imaginary parts of complex numbers; and $\circledast$ represents the convolution operator.

\section{System Model}\label{section:system}
\subsection{Near-Field Propagation Model}
\begin{figure}[t]
	\centering{}\includegraphics[scale=0.5]{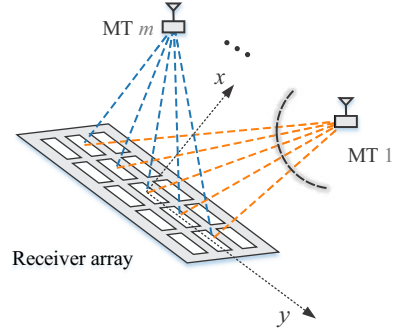}
	\caption{Near-field signals impinging at the AP equipped with an MIMO antenna.}
	\label{fig:system_model}
\end{figure}

We consider a short-range communication system composed of multiple mobile terminals (MTs) and an access point (AP) equipped with multiple antennas. As shown in Fig. \ref{fig:system_model}, each MT transmits the source signal to the AP through independent paths.
Since the dense deployment of small cells is a key feature of 5G networks, the distance between an AP and its associated terminal devices is typically no longer than 10 m \cite{7462488}. On one hand, it can provide high-quality communication services within a small area, such as offices, homes, and stadiums; on the other hand, this system of small coverage area may give rise to a high probability of line-of-sight (LoS) propagation \cite{8246850,7476821,7391158}.
Moreover, in massive MIMO scenarios, due to the large array size, it is necessary to identify the radiation pattern by calculating the Rayleigh distance $d_{\text{Ray}}=2D^2/{\lambda}$, where $D$ denotes the length or diameter of an antenna, and $\lambda$ is the carrier's wavelength. It is obvious that the large-scale antenna array tends to result in a large Rayleigh distance range.
As a result, the far-field assumption may become invalid in short-range MIMO communication scenarios. Instead, a near-field spherical wavefront model is more reasonable.

As illustrated in Fig. \ref{fig:system_model}, there are $M$ signal sources communicating with a receive-array antenna in the near-field region. The distance between the $m$-th MT and the receiver is denoted by $r_m$ ($m=1,2,\cdots,M$). The MT is located in the Fresnel zone (radiative near field) of the array when $r_m$ satisfies $0.62(D^3/ \lambda)^{1/2}< r_m <2D^2/ \lambda \label{eq:fz}$ \cite{hu2014near}. By denoting the $k$-th snapshot of the $m$-th signal as $s_m(k)$, the received signal at the $k$-th sample snapshot is written as
\begin{equation}
\mathbf{y}(k) = \mathbf{a} s_m(k)+\mathbf{z}(k),
\end{equation}
where $\mathbf{a}$ is an array steering vector and $\mathbf{z}(k)$ is a random noise vector. Note that the multipath effect is not considered since we assume the LoS propagation in the small-coverage system.

\subsection{Received Signal Model}
As far as DoA estimation is concerned, uniform linear array (ULA), uniform planar array (UPA), and uniform circular array (UCA) are widely used in traditional massive MIMO systems \cite{8377155}. Their array steering vectors are denoted by $\mathbf{a}^{\rm{ULA}}$, $\mathbf{a}^{\rm{UPA}}$, and $\mathbf{a}^{\rm{UCA}}$, respectively. The specific expressions of these array steering vectors depend on the specific antenna array geometry and the phase delay of the array elements. By introducing the interpolation method to form a virtual array, any array geometry can be transformed into a ULA for signal processing \cite{friedlander1993root}. However, this is out of the scope of the current manuscript.

In other words, the antenna arrays of arbitrary geometry are mathematically equivalent. Here and after, for ease of exposition, the ULAs are considered and let $\mathbf{a}^{\rm{ULA}} = \mathbf{a}$. We first formulate the received signals affected by the near-field effect.
To elaborate a little further, we set the center of the $N$-element ULA as the reference point (RP). Without loss of generality, when $N$ is an odd number, the $n_\text{c}$-th element (i.e., the center element) is set as the RP of the receive antenna.
In contrast to the plane-wave model, the amplitude difference of the spherical-wave signals received at different array elements cannot be ignored, thus the simple time-delay relationship among different array elements no longer holds true.

\begin{figure}[t]
	\centering{}
    \subfigure[Far-field plane wavefront.]{\includegraphics[width=2.2in]{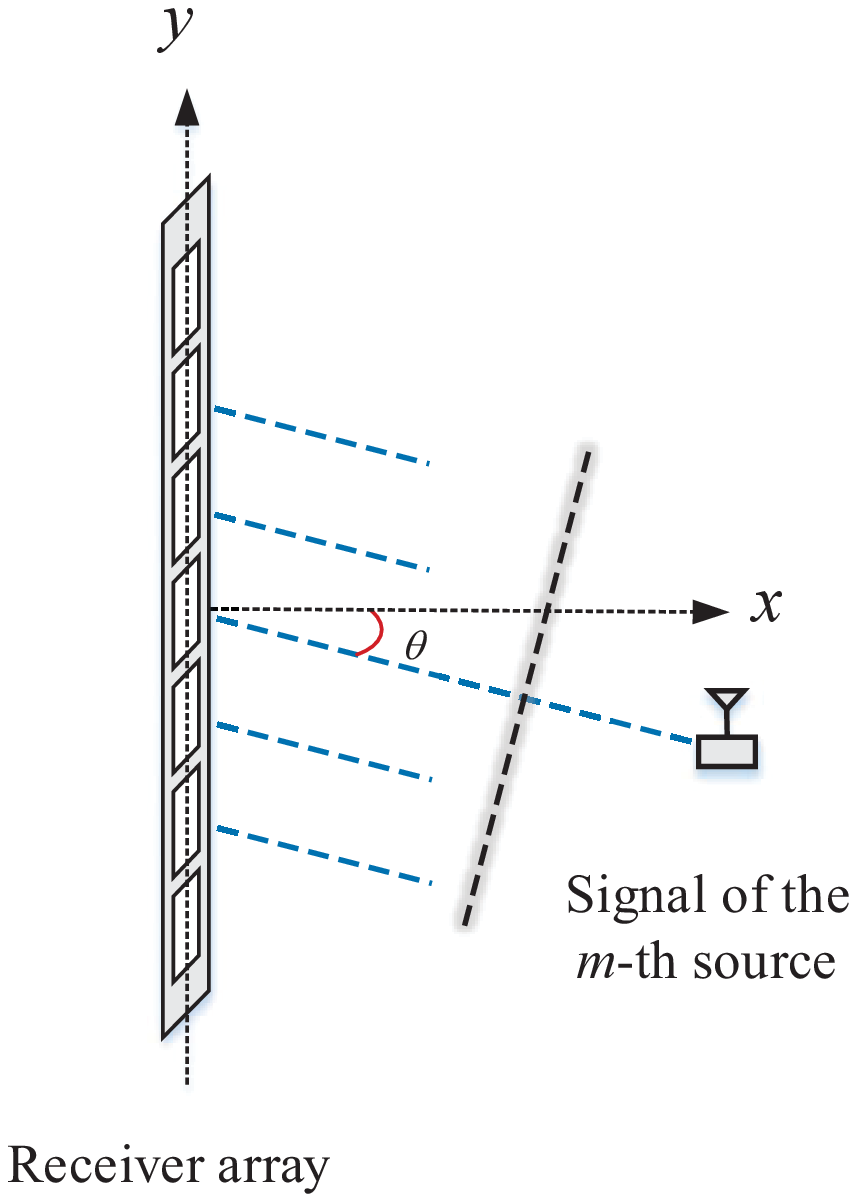}}
    \quad
    \subfigure[Near-field spherical wavefront.]{\includegraphics[width=2.2in]{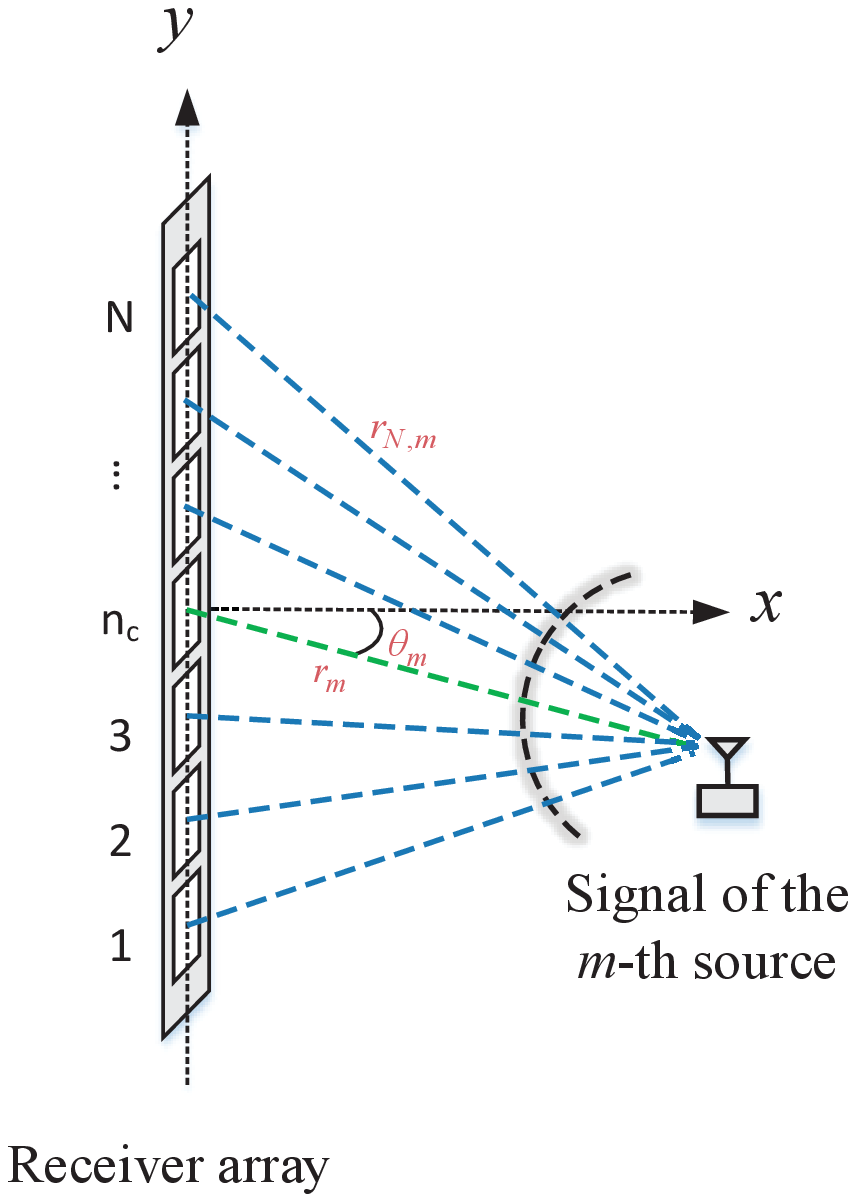}}
	\caption{The incident wave associated with the $m$-th signal source: (a) the far-field geometric propagation model with plane wavefront; (b) the near-field geometric propagation model with spherical wavefront.}
	\label{fig:signal_model}
\end{figure}

Fig. \ref{fig:signal_model}(b) intuitively illustrates the near-field geometric propagation model in a 2-D plane, where the center of the receive-array antenna is placed at the origin of the Cartesian coordinate system. Here, the $n_\text{c}$-th element is set as the RP. In this diagram, the distance from the source to the RP is denoted by $r_{{n_\text{c}},m}$, and $d$ is used to denote the element spacing. Due to the existence of resolvable paths in $M$ different directions, the model contains $M$ spherical-wavefront signals. Here, we use $\theta_m$ to represent the corresponding DoA from the $m$-th target source to the receiver, where $\theta_m \in (-{\pi}/{2}, {\pi}/{2})$.

For notational convenience, we introduce $r_{{n_\text{c}},m} = r_{m}$, and the array steering vector of the $m$-th source can be written as
\begin{align}
\mathbf{a}_{m}(\theta_{m},r_{m}) &=
\begin{bmatrix}
a_{1,m}, & \cdots & a_{n,m}, & \cdots & a_{N,m}
\end{bmatrix}^{\mathrm{T}},\label{eq:R1}
\end{align}
where
\begin{align}
a_{n,m} &= {\kappa_{n,m}}\exp \left[
-j\frac{2\pi}{\lambda}(r_{n,m}-r_{m})
\right]. \label{eq:R12}
\end{align}
In (\ref{eq:R12}), $r_{n,m}$ is the distance from the $n$-th array element to the $m$-th source, and $\kappa_{n,m} = \frac{r_{m}}{r_{n,m}}$ is the corresponding amplitude. According to the law of cosines, we have
\begin{align}
r_{n,m} = \sqrt{r_{m}^2 + \delta_n^2  d^2 - 2 r_{m} \delta_n d \sin\theta_{m}}, \label{eq:R2}
\end{align}
where $\delta_n=n-{n_\text{c}}$.
Applying the Taylor series expansion to the right-hand side of Eq. (\ref{eq:R2}), we can further get
\begin{align}
r_{n,m} & =
r_{m} - \delta_n d \sin\theta_{m} + \delta_n^2 \frac{d^2 \cos^2\theta_{m} }{ 2 r_{m} } + o \left( \frac{d^2}{r_{m}^2} \right), \nonumber\\
& = r_{m} - \delta_n d \sin\theta_{m} + \Delta_{n,m}, \label{eq:R3}
\end{align}
where $\Delta_{n,m} = \delta_n^2 \frac{d^2 \cos^2\theta_{m} }{ 2 r_{m} } + o \left( \frac{d^2}{r_{m}^2} \right)$ is the second-and higher-order Taylor polynomial term introduced by the near-field effect.
Note that, in the far-field case of planar wavefront shown in Fig. \ref{fig:signal_model}(a), we have $\kappa_{n,m}=1$ and $\Delta_{n,m}=0$.

The results of the analysis above reveal that the far-field array response model is expressed in a compact form where the phase difference between different antennas is a linear function of DoA.
The near-field array response model is characterized by a complicated function via nonlinear coupling the range-DoA parameter pair. Both amplitude and phase difference between different antennas are corrupted by range parameters. Thus, we reformulate (\ref{eq:R12}) as
\begin{align}
a_{n,m}(\theta_{m},r_{m}) &=
\frac{1}{\psi_n(\theta_{m},r_{m})}
\exp \left[
-j\frac{2\pi}{\lambda} \Phi_n(\theta_{m},r_{m})
\right], \label{eq:anp}
\end{align}
where
\begin{equation}
\psi_n(\theta_{m},r_{m}) = \frac{1}{\kappa_{n,m}} =\sqrt{
1+\left(
\frac{\delta_n d}{r_{m}}
\right)^2- \frac{2 \delta_n  d \sin\theta_{m}}{r_{m}}
 }, \label{eq:R4}
\end{equation}
and $\Phi_n(\theta_{m},r_{m})=r_m \left[\psi_n(\theta_{m},r_{m})-1\right]$.
Then, the array manifold matrix $\mathbf{A}(\boldsymbol{\theta},\mathbf{r}) \in \mathbb{C}^{N \times M}$ can be expressed as
\begin{align}
\mathbf{A}(\boldsymbol{\theta},\mathbf{r}) =
\left[
\mathbf{a}_{1}(\theta_{1},r_{1}),
\mathbf{a}_{2}(\theta_{2},r_{2}),
\cdots,
\mathbf{a}_{M}(\theta_{M},r_{M})
\right],
\end{align}
where $\boldsymbol{\theta}=[\theta_1,\cdots,\theta_M]$ and $\mathbf{r}=[r_{1},\cdots,r_{M}]$.

In practice, DoA is estimated using noisy observations. Assume that we receive $K$ snapshots of the observed signal per unit time. Therefore, the $k$-th snapshot of the complex-valued baseband signal $\mathbf{y}(k)=[y_1(k),\cdots,y_N(k)]^\mathrm{T} \in \mathbb{C}^{N}$ received at the MIMO array can be derived as
\begin{align}
\mathbf{y}(k) &=
\sum_{m=1}^{M}\mathbf{a}_{m}(\theta_{m},r_{m})s_m(k) +\mathbf{z}(k) \nonumber \\
 &=\mathbf{A}(\boldsymbol{\theta},\mathbf{r})  \mathbf{s}(k)+\mathbf{z}(k),\quad k=1,\cdots,K, \label{eq:y}
\end{align}
where $\mathbf{s}(k)=\left[s_1(k),\cdots,s_M(k)\right]^\mathrm{T}$ represents the source signal vector, and each element of $\mathbf{z}(k)=[z_1(k),\cdots,$ $z_N(k)]^\mathrm{T} \in \mathbb{C}^{N}$ follows the complex Gaussian distribution $\mathcal{CN}(0,\sigma^2)$ with zero mean and variance $\sigma^{2}$.

\section{Pre-Processing And Feature Extraction}\label{section:train}

\subsection{Problem Formulation}
For the near-field signal model shown in (\ref{eq:y}), the conventional DoA algorithm decomposes the signal covariance matrix into the noise subspace and the signal subspace. The spatial correlation matrix $\mathbf{R} \in \mathbb{C}^{N\times N}$ is given by
\begin{align}
{\bf R} = \mathbb{E}\left[\mathbf{y}(k)\mathbf{y}^\mathrm{H}(k)\right] =
\mathbf{A}\mathbf{R}_{\mathrm s} \mathbf{A}^{\mathrm{H}}+\mathbf{R}_{\mathrm z}, \label{eq:r_y}
\end{align}
where $\mathbf{R}_{\mathrm s}=\mathbb{E}\left[\mathbf{s}(k)\mathbf{s}^\mathrm{H}(k)\right]$ denotes the covariance matrix of the incident source signal, and $\mathbf{R}_{\mathrm z}=\mathbb{E}\left[\mathbf{z}(k)\mathbf{z}^\mathrm{H}(k)\right]$ represents the covariance matrix of the noise.
After executing the singular value decomposition (SVD), we rewrite Eq. (\ref{eq:r_y}) as
\begin{align}
{\mathbf R}
= \boldsymbol{\Xi}_{\mathrm s} \boldsymbol{\Lambda}_{\mathrm s} \boldsymbol{\Xi}_{\mathrm s}^{\mathrm H}+\boldsymbol{\Xi}_{\mathrm z} \boldsymbol{\Lambda}_{\mathrm z} \boldsymbol{\Xi}_{\mathrm z}^{\mathrm H}, \label{eq:usun}
\end{align}
where $\boldsymbol\Xi_{\mathrm s}$ is the signal subspace and $\boldsymbol\Xi_{\mathrm z}$ is the noise subspace; $\boldsymbol{\Lambda}_{\mathrm s}$ and $\boldsymbol{\Lambda}_{\mathrm z}$ represent the eigenvalue matrices of the signal subspace and the noise subspace, respectively. The space spanned by the array steering vectors is equivalent to that spanned by the columns of $\boldsymbol\Xi_{\mathrm s}$. In far-field DoA estimation, we can easily find ${\mathcal T}(\boldsymbol\Xi_{\mathrm s}):\boldsymbol\Xi_{\mathrm s} \mapsto \boldsymbol{\theta}$, which converts the direction finding problem into a regression problem.
As indicated in Eq. (\ref{eq:anp}), however, it should be noted that the near-field array steering vector $\mathbf{a}_m$ is the complex coupling result of two parameters $\theta_{m}$ and $r_m$. In practice, the mapping relation ${\mathcal{T}^{-1}}:\{\boldsymbol{\theta},\mathbf{r}\} \mapsto \boldsymbol\Xi_{\mathrm s}$ is revealed to be a major obstacle for learning, since tiny changes of the direct distance factor $r_m$ can result in relatively great differences of $\boldsymbol\Xi_{\mathrm s}$.
In \cite{8903003}, the authors use a DL-based scheme that exploits training datasets from different incident DoAs for the fixed $r_m$, which cannot be generalized to the signals of other distances.
Furthermore, it is impractical to acquire intensive samples with varying ranges and DoAs to build an effective dataset in the real environment. This poses a negative impact on the parameter learning task regarding distance factors with continuous intensive variables, not to mention the explosive growth of the large-scale data set. To circumvent this problem, we propose a signal preprocessing scheme, as discussed in the sequel.

\subsection{Preprocessing and Feature Selection Scheme}
The automatic feature extraction of DL methods is only effective in some specific areas. Actually, deep networks can only automatically combine and transform low-level features, so as to obtain high-level features. In digital image processing, the pixels are used as low-level feature inputs, and efficient high-level feature representations can be obtained by deep models. However, in some areas, such as natural language processing, the input word features are discrete values, which are different from continuous and dense image features. The high-level features obtained by raw data become no longer effective. Moreover, it is helpful to use \emph{a priori} knowledge to redesign features. Therefore, feature engineering is still necessary for solving DL tasks.
Due to the involvement of the parameter coupling, the direct application of neural networks to near-field DoA estimation is not straightforward. This is mainly because the coupling of parameters makes the DNN learning difficult to carry out. Here, we aim to propose an ingenious solution to preprocess the signals so that the obtained features contain only the DoA information, while eliminating the coupling effect of $r_m$.

According to Eq. (\ref{eq:y}), the received signal at each array element is given by
\begin{align}
y_n(k) = \sum_{m=1}^{M}a_{n,m}  s_m(k) + z_n(k).
\label{eq:yk}
\end{align}
Considering the fact that the far-field array manifold function has no coupling effect of range parameters, in order to facilitate analyzing the inherent far-field properties of near-field propagation, we rewrite (\ref{eq:y}) as
\begin{align}
\mathbf y(k) =
\begin{bmatrix}
\check{\mathbf a}_1^{\mathrm{H}} \\
\check{\mathbf a}_2^{\mathrm{H}} \\
\vdots \\
\check{\mathbf a}_N^{\mathrm{H}}
\end{bmatrix}\mathbf{s}(k) + \mathbf{z}(k),
\label{eq:ykn}
\end{align}
where $\check{\mathbf a}_n^{H} = \left[a_{n,1}, a_{n,2},\cdots,a_{n,M} \right]$ denotes the $n$-th row of the array manifold matrix $\mathbf{A}(\boldsymbol{\theta},\mathbf{r})$.
The entries of $\mathbf R$ can be calculated as the covariance between two array elements as follows:
\begin{align}
[\mathbf R]_{p,q} = \mathbb{E}\left[y_p(k) y_q^{\ast}(k) \right]
=\check{\mathbf a}_p^{\mathrm{H}} \mathbf{R}_{\mathrm s} \check{\mathbf a}_q + \sigma^2\delta(p-q),
\label{eq:uij}
\end{align}
where $p,q \in \{1,2,\cdots,N \}$ and $\delta(\cdot)$ is the Kronecker delta function.

Recalling (\ref{eq:R3}) and (\ref{eq:anp}), we find that by the Taylor series $\Phi_n(\theta_{m},r_{m})$ can be expressed in a tractable Fresnel approximation \cite{hu2014near} form as
\begin{align}
\Phi_n(\theta_{m},r_{m}) = -\delta_n d\sin\theta_{m} + \delta_n^2 \frac{d^2 \cos^2\theta_{m}}{2 r_{m}}. \label{eq:Rf}
\end{align}
Substituting (\ref{eq:Rf}) into (\ref{eq:anp}) yields
\begin{align}
a_{n,m} &= \kappa_{n,m} e^{ j \left( \delta_n \frac{2\pi d}{\lambda}\sin\theta_{m}
- \delta_n^2 \frac{\pi d^2}{\lambda r_{m}}\cos^2\theta_{m} \right) }.
\label{eq:a_f}
\end{align}
Then, (\ref{eq:yk}) can be directly reframed as
\begin{align}
y_n(k) = \sum_{m=1}^{M} \kappa_{n,m} e^{ j\left(
\delta_n \alpha_m -\delta_n^2 \beta_m
\right)  }
 s_m(k) + z_n(k),
\label{eq:yf}
\end{align}
where $\alpha_m = \frac{2\pi d}{\lambda}\sin\theta_{m}$ and $\beta_m = \frac{\pi d^2}{\lambda r_{m}}\cos^2\theta_{m}$.

It should be noted that $\beta_m$ is corrupted by the distance factor $r_{m}$, while $\alpha_m$ is not corrupted. We know that $\alpha_m$ reflects the phase delay of the far-field mode. The associated far-field array steering vectors are given by
\begin{equation}
\mathbf{a}_m(\theta_m) = \left[ 1, e^{j\frac{2\pi d}{\lambda}\sin\theta_{m}},\cdots, e^{j\frac{2\pi (N-1)d}{\lambda}\sin\theta_{m}} \right]^{\mathrm{T}}. \label{eq:ffa_vec}
\end{equation}

Substituting (\ref{eq:yf}) into (\ref{eq:uij}), we have
\begin{align}
[\mathbf R]_{p,q} =&
\sum_{m=1}^{M} \kappa_{p,m} \kappa_{q,m} [\mathbf{R}_{\mathrm s}]_{m,m}e^{ j
(p-q) \alpha_m} \nonumber \\
&e^{ -j(\delta_p^2-\delta_q^2)\beta_m } + \sigma^2\delta(p-q),\label{eq:ffa1}
\end{align}
where $\delta_p = p-n_\text{c}$ and $\delta_p \in \{\frac{1-N}{2}, \cdots, 0, \cdots, \frac{N-1}{2}\}$. From Eq. (\ref{eq:ffa1}), we can obtain the corresponding expression of far-field sources with identical incident DoAs group as follows:
\begin{align}
[\bar{\mathbf R}]_{p,q} = \sum_{m=1}^{M} [\mathbf{R}_{\mathrm s}]_{m,m} e^{j
(p-q) \alpha_m  } + \sigma^2\delta(p-q),\label{eq:ffa2}
\end{align}
where $\bar{\mathbf R}$ indicates the ideal far-field signal covariance matrix.
By comparing (\ref{eq:ffa1}) and (\ref{eq:ffa2}), the distinguishing amplitude feature for the $m$-th source between far field and near field can be calculated as
\begin{align}
\kappa_{p,m} \kappa_{q,m} \approx 1 + \frac{d \sin\theta_m}{r_m}(\delta_p+\delta_q).
\label{eq:amp_f}
\end{align}
For $q=p+t$ where $t \in \{-(N-1),-(N-2),\cdots,N-1\}$, the squared magnitude of the element-wise differences in the far-field and near-field patterns is given by (\ref{eq:ffa3}) (at the top of the next page),
\begin{figure*}[t]
\begin{align}
\bigg\vert [\mathbf R]_{p,p+t} - [\bar{\mathbf R}]_{p,p+t} \bigg\vert^2
\approx \sum_{m=1}^{M} \Big\vert[\mathbf{R}_{\mathrm s}]_{m,m}\Big\vert^2 \left[
\left( \left(t+2\delta_p\right) \frac{d \sin\theta_m}{r_m}  \right)^2 +
\left( \kappa_{p,m} \kappa_{q,m} t\left(t+2\delta_p\right)\frac{\pi d^2}{\lambda r_m}\cos^2 \theta_m\right)^2 \right].
\label{eq:ffa3}
\end{align}
\hrulefill
\end{figure*}
which represents the approximation error.
Then, for the purpose of minimizing the approximation error in (\ref{eq:ffa3}), a VCM approximating the far-field covariance matrix can be constructed according to the Hermitian and Toeplitz properties of the far-field covariance matrix.
It is easily seen that if $t/2$ is an integer, (\ref{eq:ffa3}) achieves the minimum error at $\text{x}_p=-t/2$; otherwise, if $(t-1)/2$ is an integer, (\ref{eq:ffa3}) reaches the minimum value at $\delta_p=-(t-1)/2$ or $\delta_p=-(t+1)/2$.
In light of the above observation, the VCM is constructed as $\widehat{\mathbf R}$ according to
\begin{align}
[\widehat{\mathbf R}]_{p,p+t} = \frac{[\mathbf R]_{\chi_l(t),\chi_l(-t)} + [\mathbf R]_{\chi_r(t),\chi_r(-t)}}{2},
\label{eq:ffa4}
\end{align}
where $\chi_l(t) = \lfloor n_\text{c}-\frac{t}{2} \rfloor$ and $\chi_r(t) = \lfloor n_\text{c}-\frac{t-1}{2} \rfloor$.

\begin{figure}[t]
	\centering{}\includegraphics[scale=0.5]{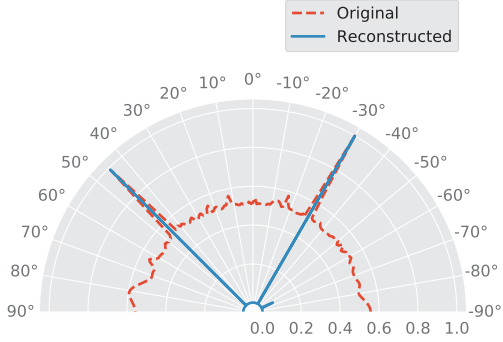}
	\caption{Beam patterns of original and reconstructed covariance matrices with two signals, which are propagated from $-30^{\circ}$ and $45^{\circ}$.}
	\label{fig:err}
\end{figure}

Fig. \ref{fig:err} shows the beam patterns obtained based on $\mathbf R$ and $\widehat{\mathbf R}$, respectively. We can see that, compared with the original covariance matrix based scheme, the reconstructed VCM based scheme exhibits sharper peaks at the DoAs and lower sidelobes. Applying SVD to the reconstructed VCM (\ref{eq:ffa4}), we readily obtain
\begin{align}
\widehat{\mathbf R} = \widehat{\boldsymbol\Xi}_{\mathrm s} \widehat{\boldsymbol\Lambda}_{\mathrm s} \widehat{\boldsymbol\Xi}_{\mathrm s}^{\mathrm H}+\widehat{\boldsymbol\Xi}_{\mathrm z} \widehat{\boldsymbol\Lambda}_{\mathrm z} \widehat{\boldsymbol\Xi}_{\mathrm z}^{\mathrm H},\label{eq:svd}
\end{align}
where $\widehat{\boldsymbol\Xi}_{\mathrm s}=\left[\boldsymbol\xi_1,\boldsymbol\xi_2,\cdots,\boldsymbol\xi_M\right] \in \mathbb{C}^{N\times M}$ is the eigenvector matrix of the signal subspace and $\widehat{\boldsymbol\Xi}_{\mathrm z}$ is the eigenvector matrix of the noise subspace; $\widehat{\boldsymbol{\Lambda}}_{\mathrm s}$ and $\widehat{\boldsymbol{\Lambda}}_{\mathrm z}$ denote the eigenvalue matrices of the signal subspace and the noise subspace, respectively.
In far-field array signal processing, the subspace spanned by the far-field array steering vectors is equal to the eigenvector matrix $\widehat{\boldsymbol\Xi}_{\mathrm s}$, i.e.,
\begin{align}
\text{span} \left[
\boldsymbol\xi_1,\cdots,\boldsymbol\xi_M
\right] =
\text{span} \left[
\mathbf{a}_1(\theta_{1}),\cdots,\mathbf{a}_M(\theta_{M})
\right]
,\label{eq:span}
\end{align}
where $\text{span}[\cdot]$ stands for the span of a set of vectors.

\section{Proposed Deep Complex Network Architecture}\label{section:cvnn}

\subsection{Deep Complex Network Architecture}
In signal processing, for the convenience of signal representation, we usually deal with complex-valued signals that contain both the amplitude and phase information. Inspired by \cite{zhang2017complex} and \cite{trabelsi2017deep}, we design a deep complex network for meaningful information extraction, which is used to carry out the DoA estimation.

The proposed complex-valued network is able to simulate the complex arithmetic operations in an internal mechanism.
Let $c$ denote a complex number in the form of $c=u+j v$ with $u,v \in \mathbb{R}$.
Assume that the input of the $l$-th layer of the network is a complex vector $\mathbf{c}^{l}=\mathbf{u}^{l}+j \mathbf{v}^{l}$, where $\mathbf{u}^{l}$ and $\mathbf{v}^{l}$ are real vectors.
In the $l$-th layer of the proposed CVNN model, the complex weight matrix $\mathbf{W}^l$ and complex bias vector $\mathbf{b}^l$ are expressed as
\begin{align}
\mathbf{W}^l=\mathbf{W}^l_{\mathrm{R}} + j\mathbf{W}^l_{\mathrm{I}}, \quad\quad
\mathbf{b}^l= \mathbf{b}^l_{\mathrm{R}} + j\mathbf{b}^l_{\mathrm{I}},
\end{align}
where $\mathbf{W}^l_{\mathrm{R}}$ and $\mathbf{W}^l_{\mathrm{I}}$ correspond to the real and imaginary parts of $\mathbf{W}^l$, respectively; $\mathbf{b}^l_{\mathrm{R}}$ and $\mathbf{b}^l_{\mathrm{I}}$ denote the real and imaginary parts of $\mathbf{b}^l$, respectively.

\begin{figure}[t]
	\centering{}\includegraphics[scale=0.2]{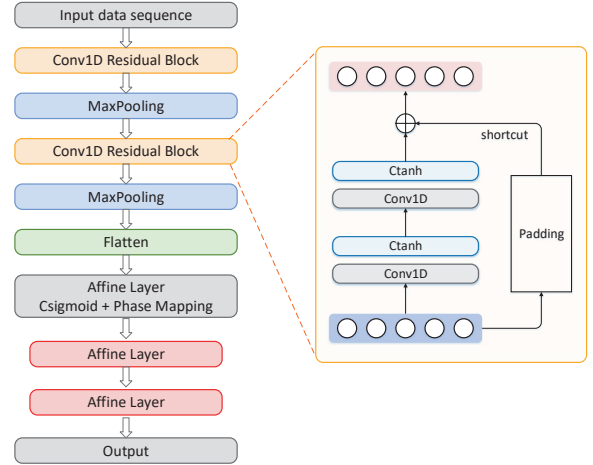}
	\caption{The deep complex ResNet model. The network is shown on the left side, and the right side represents the Conv1D residual block diagram.}
	\label{fig:stru}
\end{figure}

Empirically, the depth of the network plays an important role in classification or regression tasks from a training point of view. However, deeper networks may result in more intractable issues, such as the gradient exploding or vanishing.
Inspired by the ResNet proposed in \cite{he2016deep}, we develop a residual block architecture consisting of several blocks, where a shortcut connection is added to realize the linear superposition of the input itself and its nonlinear transformation in each block.
The specific components of each block are shown in Fig. \ref{fig:stru}.

As shown in Fig. \ref{fig:stru}, the architecture of the proposed complex ResNet is composed of two alternations of convolutional and pooling layers, a flatten layer, and three affine layers. In our proposed deep complex network, two 1-D convolution (Conv1D) layers are contained in a convolutional residual block. We mainly employ 2 residual blocks to learn the phase-related features of the input complex-valued data. The residual block is followed by a pooling layer, which extracts the critical features.
The results after pooling are flattened into a column and fed into a complex affine layer, whose outputs are mapped to real-valued phases by specially devised activation functions.
Finally, since we define the near-field DoA estimation as a regression problem, several real affine layers must be connected to the end of network to predict the continuous phase variable.
The details of each type of layer representation are listed in the following.

\begin{figure}[t]
	\centering{}\includegraphics[scale=0.3]{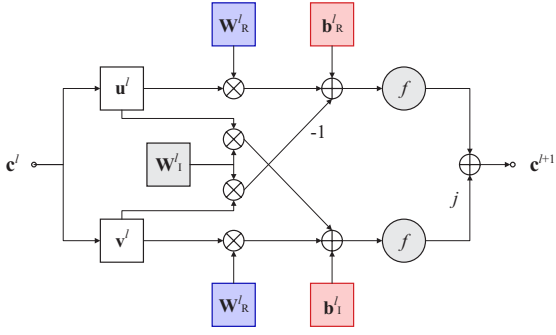}
	\caption{The schematic model of the affine layer of the proposed CVNN.}
	\label{fig:layer}
\end{figure}

1) Affine layer: Affine means that the neurons between adjacent layers are fully connected. The complex-valued arithmetic process is expressed as
\begin{align}
\begin{bmatrix}
\Re(\mathbf{W}^l\mathbf{c}^l) \\
\Im(\mathbf{W}^l\mathbf{c}^l)
\end{bmatrix} =
\begin{bmatrix}
\mathbf{W}_{\mathrm{R}}^l & -\mathbf{W}_{\mathrm{I}}^l\\
\mathbf{W}_{\mathrm{I}}^l & \mathbf{W}_{\mathrm{R}}^l
\end{bmatrix}
\begin{bmatrix}
\mathbf{u}^l\\
\mathbf{v}^l
\end{bmatrix}.
\label{eq:fc}
\end{align}
The signal flow diagram of the complex-valued affine layer is shown in Fig. \ref{fig:layer}.

2) Convolution layer: The convolution operation of the CVNN, analogous to affine layer, is expressed as
\begin{align}
\mathbf{W}^l\circledast\mathbf{c}^l =& (\mathbf{W}_{\mathrm{R}}^l\circledast\mathbf{u}^l-\mathbf{W}_{\mathrm{I}}^l\circledast\mathbf{v}^l) \nonumber\\
& + j(\mathbf{W}_{\mathrm{I}}^l\circledast\mathbf{u}^l + \mathbf{W}_{\mathrm{R}}^l\circledast\mathbf{v}^l).
\label{eq:co}
\end{align}
Additionally, the Conv1D filters are employed in 1-D input features analysis.
If the size of the input matrix is $L_i\times F_i$, where $L_i$ is the length of the input vector and $F_i$ is the number of the feature channels, then the required filter has a dimension of $L_c\times F_i \times F_c\times S_c$, where $S_c$ is the stride used to describe the step size of the convolution operation and $L_c$ is the length of the filter.
In addition, $F_c$ represents the number of filters, and it determines the number of output channels.
A filling-up convolution method, known as ``SAME'' padding, is conducted to avoid losing information from the original data.
Then, the dimension of the output after Conv1D will be $\lceil L_i/S_c \rceil \times F_c = L_{co}\times F_c$ for ``SAME'' padding. After applying the 1-D pooling layer with the $L_p\times S_p$ kernel, the final size of the output becomes $\lceil L_{co}/S_p \rceil \times F_c$, where $L_p$ denotes the length of the pooling kernel and $S_p$ is the stride of the pooling kernel.

3) Pooling layer: In convolutional neural networks (CNNs), the pooling layer is often placed behind the convolution layer to incorporate the features from small neighborhoods. As a result, the key features for prediction can be obtained, while sparse and valid information can be retained. In this paper, the max-pooling layers are applied to the real and imaginary parts, respectively.

\subsection{Activation Function and Backpropagation}
For CVNNs, the activation function is the most critical part, since it directly determines the nonlinear fitting ability and the convergence of the back propagation (BP) algorithms \cite{huang2018regression}.
Generally speaking, complex activation functions should be analytic and bounded. In other words, the functions should satisfy the Cauchy-Riemann (C-R) equation.
In the complex analysis, Liouville's theorem states that complex functions that are both analytic and bounded must be constant functions \cite{scardapane2018complex}. However, in existing literature, many researchers prefer bounded functions to analytic functions.
Simultaneously, due to the fact that the phase information is determined by both the real and the imaginary parts, we consider an activation function that affects the magnitude to indirectly achieve the phase learning. To this end, a split complex function is employed in convolution layers, and it is given by
\begin{align}
\mathrm{Ctanh}(c) = \tanh(u)+j\tanh(v).
\end{align}
Although the activation function $f(c) = \mathrm{Ctanh}(c)$ is non-analytic, it is bounded in the complex domain. The boundedness of the activation function eliminates the appearance of singular points.
Considering that our training targets have values in the interval $(-\pi/2,\pi/2)$, it will cause a large angle error in degrees even when the small training loss is reached. Rescaling from the target values in radians does not work. Note that the mean of the target data is zero. To make the model focus more on small differences, we adopt the tanh function in affine layers since it has the highest slope at the zero point. Such a design can cause the gradient boosting with small prediction differences.

We also devise a special activation function for a mapping from the complex domain to the real domain, thus facilitating the measurement of prediction errors. To be specific, after the complex affine layer we define the following activation functions:
\begin{align}
\mathrm{Csigmoid}(c) &= \text{sigmoid}(u)+j \text{sigmoid}(v), \label{eq:sigmoid} \\
\rho &= \arctan \frac{\Im(c)}{\Re(c)} =\arctan\left(\frac{c-{c}^{\ast}}{c+{c}^{\ast}}\right), \label{eq:phase}
\end{align}
where $\rho$ is the real-valued phase corresponding to the output of complex affine layer.
To avoid the invalid division in (\ref{eq:phase}), we employ the sigmoid function in (\ref{eq:sigmoid}) to keep the real part of the output greater than 0.

In CVNNs, the derivable function $\mathcal L$ regarding the real and imaginary parts of parameters can also be applied to BP, respectively \cite{6138313}.
The derivative processing for the real and imaginary parts is written as
\begin{align}
\frac{\partial \mathcal L}{\partial \mathbf{c}}
&= \Re(\nabla \mathcal L(\mathbf{c}))+j\Im(\nabla \mathcal L(\mathbf{c})),\label{eq:chain_1}\\
\frac{\partial \mathcal L}{\partial \mathbf{W}}
&=\Re(\nabla \mathcal L(\mathbf{c}))\left(\frac{\partial \mathbf{u}}{\partial \mathbf{W}_\mathrm{R}}+j\frac{\partial \mathbf{u}}{\partial \mathbf{W}_\mathrm{I}}\right)  + \Im(\nabla \mathcal L(\mathbf{c}))\nonumber\\
&\left(\frac{\partial \mathbf{v}}{\partial \mathbf{W}_\mathrm{R}}+j\frac{\partial \mathbf{v}}{\partial \mathbf{W}_\mathrm{I}}\right).
\label{eq:chain_2}
\end{align}
Accordingly, the network parameters $\mathbf{W}^{l}_{\mathrm{R}}$ and $\mathbf{W}^{l}_{\mathrm{I}}$ are iteratively updated by means of the stochastic gradient descent (SGD) algorithm \cite{huang2018regression}.

\subsection{Training Strategy}

\begin{figure}[t]
    \centering{}\includegraphics[scale=0.8]{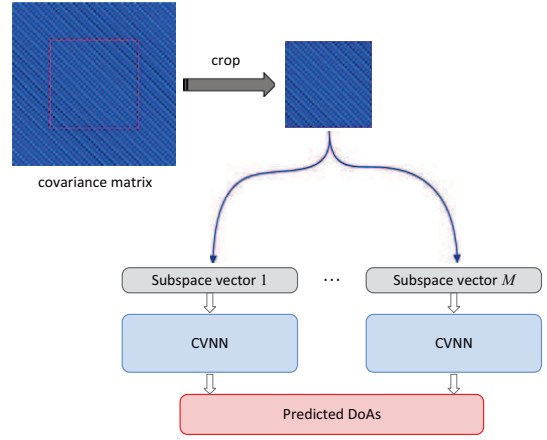}
	\caption{The CVDL based DoA estimation framework.}
	\label{fig:framework}
\end{figure}

We intend to propose a strategy which can be applied to different numbers of antennas at the receiver array, instead of training special DNNs with different input dimensions corresponding to various numbers of antenna elements. Different from the existing methods, we design new input features for a general deep network in our DoA estimation framework. The training strategy can be seen from Fig. \ref{fig:framework}.
Above all, the VCM derived from any $N$-element antenna is cropped to a fixed size so as to retain the key features.
The cropped VCM is constructed as
\begin{align}
\underline{\mathbf R} &= \left[\widehat{\mathbf R} \right]_{\frac{N-N_{\text{in}}}{2}+1:N-\frac{N-N_{\text{in}}}{2}+1, \frac{N-N_{\text{in}}}{2}+1:N-\frac{N-N_{\text{in}}}{2}+1} \nonumber\\
&= \underline{\boldsymbol\Xi}_{\mathrm s} \underline{\boldsymbol\Lambda}_{\mathrm s} \underline{\boldsymbol\Xi}_{\mathrm s}^{\mathrm H}+\underline{\boldsymbol\Xi}_{\mathrm z} \underline{\boldsymbol\Lambda}_{\mathrm z} \underline{\boldsymbol\Xi}_{\mathrm z}^{\mathrm H},
\label{eq:crop}
\end{align}
where $N_{\text{in}} \times N_{\text{in}} (N_{\text{in}} < N)$ is the cropped block size of the original VCM.
Given the signal subspace vectors $\underline{\mathbf \Xi}_{\mathrm s} = \left[\underline{\boldsymbol\xi}_1, \cdots, \underline{\boldsymbol\xi}_M \right]$, the input features are constructed as
\begin{align}
\mathbf{c}_{\text{input}} = \underline{\boldsymbol\xi}_m \in \mathbb{C}^{N_{\text{in}}}, \quad m=1,2,\cdots,M.
\label{eq:input}
\end{align}
Finally, independent subspace vectors are adopted as training samples to perform the learning task.

\begin{algorithm}[htbp]
\small
\caption{The proposed CVDL-aided DoA estimation method}
\label{alg:A1}
\begin{algorithmic}[1]
\REQUIRE ~~\\ 
The received signals $\mathbf{y}(k),\quad k=1,2,\cdots,K$;\\
The number of near-field sources $M$;\\
\ENSURE ~~\\ 
The estimates of DoA, $\hat{\boldsymbol{\theta}}$;
\STATE Calculate the covariance matrix $\bf R$ based on the samples $\mathbf{y}(k)$ according to (\ref{eq:r_y});
\label{ code:fram:r_y }
\STATE Reconstruct the VCM $\widehat{\bf R}$ according to
$$[\widehat{\mathbf R}]_{p,p+t} = \{[\mathbf R]_{\delta_l(t),\delta_l(-t)} + [\mathbf R]_{\chi_r(t),\chi_r(-t)}\}/{2};$$
\label{code:fram:anti}
\STATE Crop the VCM $\widehat{\bf R}$ obtained by (\ref{eq:ffa4}) to acquire the compressed signal subspace $\underline{\boldsymbol{\Xi}}_{\mathrm{s}}$ based on (\ref{eq:crop});
\label{code:fram:tp}
\STATE Devise the input feature representations $\mathbf{c}_{\text{input}}$ based on (\ref{eq:input}) to build the training samples;
\label{code:fram:input}
\STATE Feed the data into the proposed CVNN depicted in Fig. \ref{fig:stru}, and perform forward propagation to obtain the network output $\hat{\boldsymbol{\theta}}$;
\label{code:fram:cvnn}
\RETURN $\hat{\boldsymbol{\theta}}$; 
\end{algorithmic}
\end{algorithm}

In this paper, more attention is paid to angle prediction in regression tasks.
Both mean squared error (MSE) and mean absolute error (MAE) are popularly used for regression loss metrics \cite{ghosh2017robust}. Although the MSE loss is easy to compute the gradient, large errors have relatively greater influence on the MSE than the small errors due to the square.
Compared with the MSE, the MAE loss is more robust to outliers since it does not use square. Moreover, the MAE is more useful if we are concerned about small errors such as subtle radian errors.
Therefore, a typical MAE loss function is defined as
\begin{align}
\mathcal L (\hat{{\boldsymbol\theta}}, \boldsymbol\theta)
&= \frac{1}{BM} \sum_{b=1}^{B} \Vert \hat{{\boldsymbol\theta}}_b - \boldsymbol\theta_b \Vert_1, \label{eq:mae}
\end{align}
where $B$ is the total number of data samples. The true and estimated DoAs for the $b$-th sample are denoted by ${\boldsymbol\theta}_b$ and $\hat{\boldsymbol\theta}_b$, respectively.

For the purpose of extending to multi-source DoA estimation, we employ a training strategy that only uses the signal subspace vector as training samples. As such, the subspace vectors associated with multi-DoA are fed into the network separately to train and predict. The proposed CVDL-aided DoA estimation method is summarised in Algorithm \ref{alg:A1}.

\subsection{Complexity Analysis}
We use the total number of floating-point operations (FLOPs) to measure the computational complexity of the proposed network. The computational complexities of both the convolutional and dense layer operations are 4 times more expensive than their real counterparts but with the same order \cite{trabelsi2017deep}.
According to \cite{MolchanovTKAK17}, in real-valued networks, the FLOPs of a conv1D layer is given by $2L_i(F_i L_c^2+1)F_c$ and the FLOPs of a dense layer is given by $(2I-1)O$, where $I$ and $O$ denote the input and output dimensions, respectively.

\begin{table}[t]
	\centering\small
	\caption{CVNN Hyperparameters}
	\label{tb:t_nn}
	\begin{spacing}{1.2}
	\begin{tabular}{c|c|c}
		\hline\hline
		\multicolumn{2}{c|}{\textbf{Layer Name}}              & \multicolumn{1}{c}{\textbf{Configuration}} \\ \hline\hline
		\multicolumn{2}{c|}{Input} & $N_{\text{in}}\times 1$  \\ \hline
		\multirow{2}{*}{Residual Block 1}
        & Conv1D $\times 2$ & $3\times1\times8\times1$                           \\ \cline{2-3}
		& Conv1D $\times 2$ & $3\times8\times8\times1$                           \\ \hline
		\multicolumn{2}{c|}{MaxPooling}              & $2\times2$                \\ \hline
		\multirow{2}{*}{Residual Block 2}
        & Conv1D $\times 2$ & $3\times8\times4\times1$                           \\ \cline{2-3}
		& Conv1D $\times 2$ & $3\times4\times4\times1$                           \\ \hline
		\multicolumn{2}{c|}{MaxPooling}              & $2\times2$                \\ \hline
        \multicolumn{2}{c|}{Flatten}                 & $N_{\text{flat}}$        \\ \hline
        \multicolumn{2}{c|}{Affine (Csigmoid + Phase Mapping)}   & $20$         \\ \hline
        \multicolumn{2}{c|}{Affine}           & $10$         \\ \hline
		\multicolumn{2}{c|}{Affine}                  & $10$                      \\ \hline
		\multicolumn{2}{c|}{Output}  & $1$                      \\ \hline \hline
	\end{tabular}
    \end{spacing}
\end{table}

\begin{table}[t]
	\centering\small
	\caption{RVNN (TDNN \cite{huang2018regression}) Architecture}
	\label{tb:t_dnn}
	\begin{spacing}{1.2}
    \begin{tabular}{c|c}
        \hline\hline
        Layer Name     & Layer Specification    \\ \hline\hline
        input layer    & $\small{\texttt{input dim}}=N_{\text{in}}'$    \\ \hline
        hidden layer 1 & $\small{\texttt{context size}}=5$,  $\small{\texttt{filter num}}=8$   \\ \hline
        hidden layer 2 & $\small{\texttt{context size}}=5$,  $\small{\texttt{filter num}}=8$   \\ \hline
        hidden layer 3 & $\small{\texttt{context size}}=5$,  $\small{\texttt{filter num}}=4$   \\ \hline
        hidden layer 4 & $\small{\texttt{context size}}=5$,  $\small{\texttt{filter num}}=2$   \\ \hline
        hidden layer 5 & $\small{\texttt{context size}}=5$,  $\small{\texttt{filter num}}=1$   \\ \hline
        dense layer    & $\small{\texttt{output dim}}=10$                  \\ \hline
        dense layer    & $\small{\texttt{output dim}}=10$                  \\ \hline
        output layer   & $\small{\texttt{output dim}}=1$                   \\ \hline\hline
    \end{tabular}
    \end{spacing}
\end{table}

Regarding the deep complex model, the hyperparameters of the CVNN are detailed in Table \ref{tb:t_nn}. These parameters are listed according to the order of the layers depicted in Fig. \ref{fig:stru}, and the network configuration $L_c\times F_i \times F_c\times S_c$ corresponds to the numerical expression of the Conv1D kernels in the third column of Table \ref{tb:t_nn}.
For the training and test sets, each sample of the dataset contains $N_{\text{in}}$ entries taken from the signal subspace $\underline{\mathbf \Xi}_{\mathrm s}$ generated by using (\ref{eq:crop}). The number of neurons in the flatten layer is denoted as $N_{\text{flat}}$.

The time delay neural network (TDNN) architecture proposed in \cite{huang2018regression} has been taken as the RVNN-based baseline scheme for regression prediction, and its structure and parameter settings are shown in Table \ref{tb:t_dnn}. Since the RVNN cannot directly deal with the complex-valued features, according to \cite{huang2018regression}, the real and imaginary parts of (\ref{eq:input}) are concatenated as the input tensor, i.e.,
\begin{align}
\mathbf{u}_{\text{input}}=
\begin{bmatrix}
\Re(\mathbf{c}_{\text{input}}) \\
\Im(\mathbf{c}_{\text{input}})
\end{bmatrix}
 \in \mathbb{R}^{N_{\text{in}}'},
\label{eq:dnnin}
\end{align}
where $N_{\text{in}}' = 2\times N_{\text{in}}$ indicates that the input size of the RVNN is twice as the input size of the CVNN.

For the case of $N_{\text{in}}=33$, the total number of FLOPs of the proposed CVNN is about 0.24 million. In this case, we have $N_{\text{in}}'=66$ for the real TDNN and its total number of FLOPs is about 0.34 million. Note that the proposed CVNN can keep operations within complex structures while conventional RVNNs achieve brute-force fitting by increasing network depth and dimensions.

\begin{table}[ht]
	\centering\small
	\caption{Parameters of the Near-Field Source Dataset}
	\label{tb:source}
	\begin{spacing}{1.2}
		\begin{tabular}{c|c|c}
			\hline\hline
			\multirow{2}{*}{Parameters} &
			\multicolumn{2}{c}{Value and Range} \\
            \cline{2-3}
			& Training Set   & Test Set \\ \hline\hline
			Distance (in $\lambda$) & (200, 1800) step: 25     & (200, 1800) step: 25 \\
            \cline{1-3}
			Direction (${}^{\circ}$) & (-90, 90) step: 0.01 & (-90, 90) step: 0.1 \\
			\cline{1-3}
			Snapshots   & 100    &  100   \\
			\cline{1-3}
            SNR (dB)    & 10    &  10 \\ \hline\hline
		\end{tabular}
	\end{spacing}
\end{table}

\begin{figure}[t]
    \centering
    \subfigure[]{\includegraphics[scale=0.35]{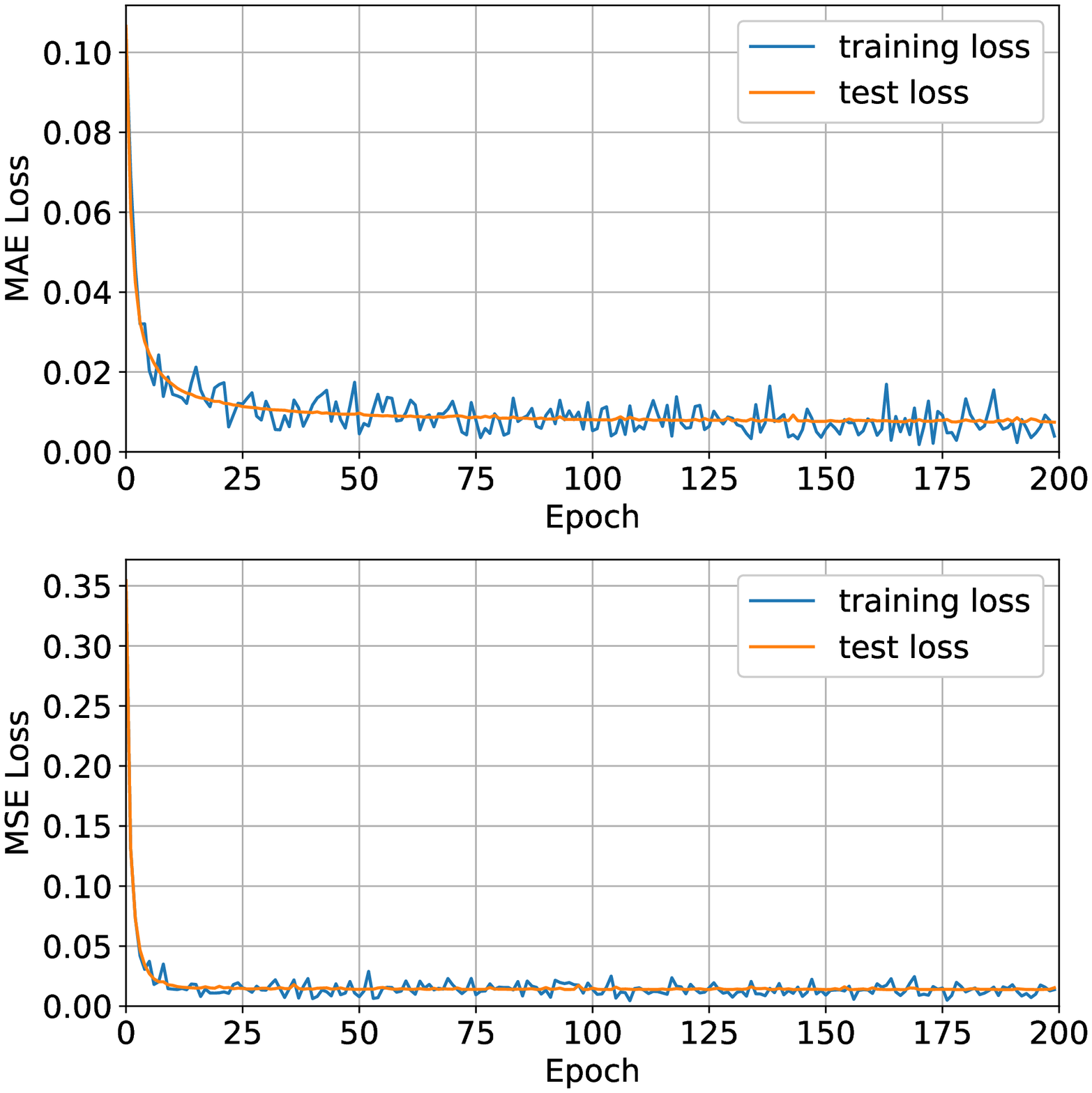}}
    \quad
    \subfigure[]{\includegraphics[scale=0.35]{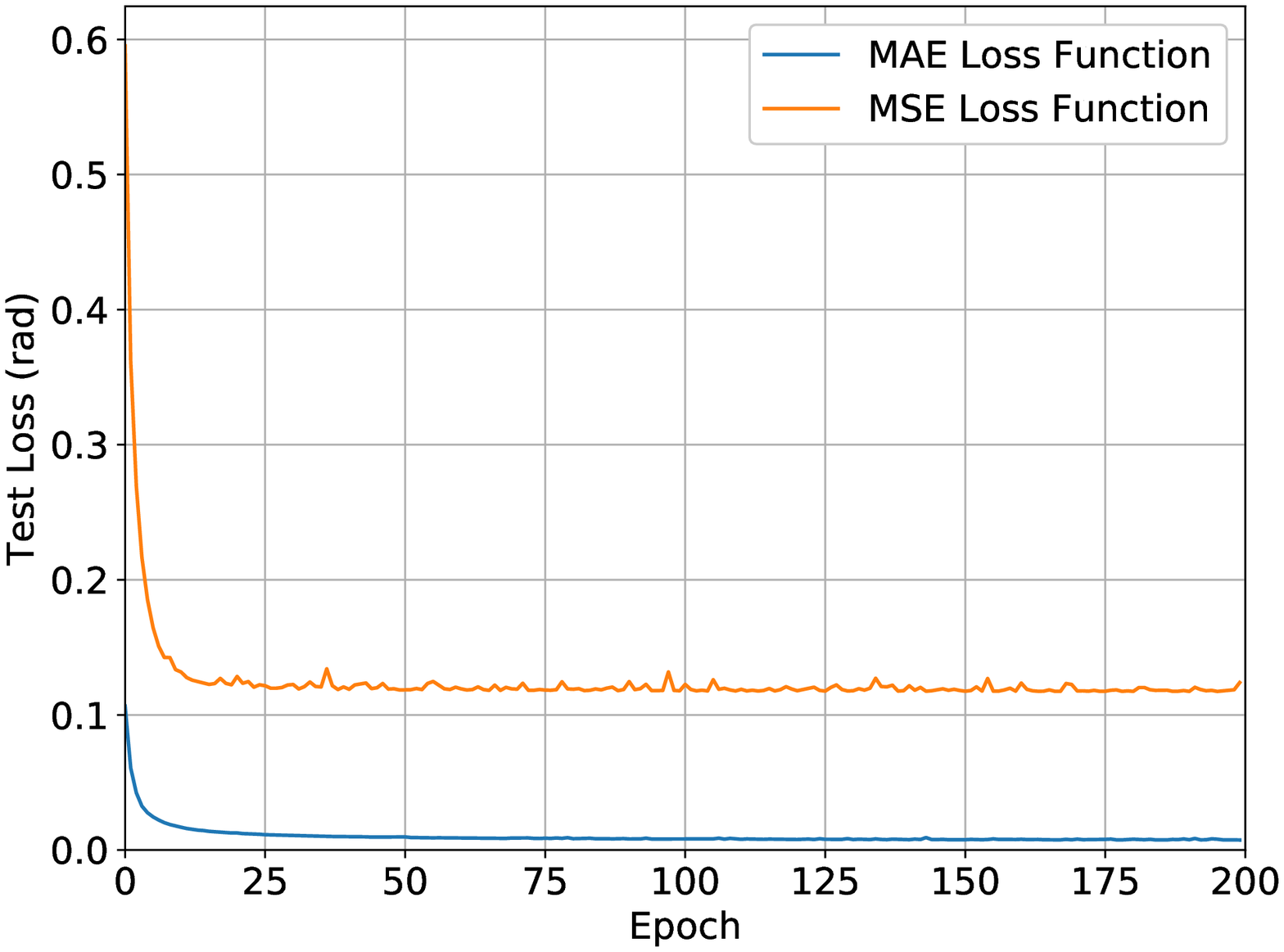}}
    \caption{Model training results for different loss functions. (a) Learning curve of 200 epochs in the offline training phase. (b) Comparing test errors trained with different loss metrics.}
    \label{fig:loss}
\end{figure}

\section{Simulation Results and Discussions}\label{section:sim}
\subsection{Simulation Setup}

In this section, the CVDL-aided DoA estimation approach is evaluated by numerous simulations. All parameter settings are in accordance with the near-field region. The implementation of our deep network is based on TensorFlow \cite{abadi2016tensorflow}.
The estimation results are compared with the MUSIC method in \cite{6509484}, the SVR method in \cite{7848281} and the real TDNN scheme in \cite{huang2018regression}.
The root mean square error (RMSE) is used to evaluate our proposed model.

To elaborate a little further, the transmitted signals with carrier frequency of $f_c = 28$ GHz are used. The corresponding wavelength is $\lambda = 0.0107$ m.
We consider a ULA with $N=65$ array elements, thus the position of the reference array element is $n_\text{c}=33$. We set the array spacing as $d=\lambda/2$. Then, the array aperture is $D=32\lambda$, and the Fresnel region is between $112\lambda$ and $2048\lambda$ \cite{hu2014near}. 

For the dataset, the received signals are collected by using (\ref{eq:y}), and the input datasets are generated according to the scheme proposed in Section \ref{section:train}-B.
In attempting to capture the impact of multiple factors, the dataset should cover a variety of conditions, such as snapshots, directions and distance from the reference array element, which are summarized in Table \ref{tb:source}.
Note that we train the CVNN offline by employing the raw data with additive noise as the training samples, where the SNR and the number of snapshots are fixed at 10 dB and 100, respectively. Then we will test the performance of online deployment under various SNRs and snapshots.
Based on our CVNN framework, the learning model of multi-source scenarios can be trained via adding subspace vector samples associated with multi-DoA. As an initial work and for the sake of simplicity, we take $M=1$ as an example.

\begin{figure}[ht]
    \centering\includegraphics[scale=0.45]{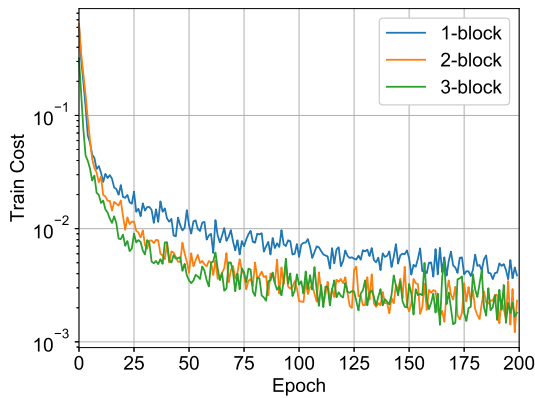}
    \caption{Residual block's impact on learning.}
    \label{fig:block}
\end{figure}

\begin{figure*}[t]
    \centering{}\includegraphics[scale=0.6]{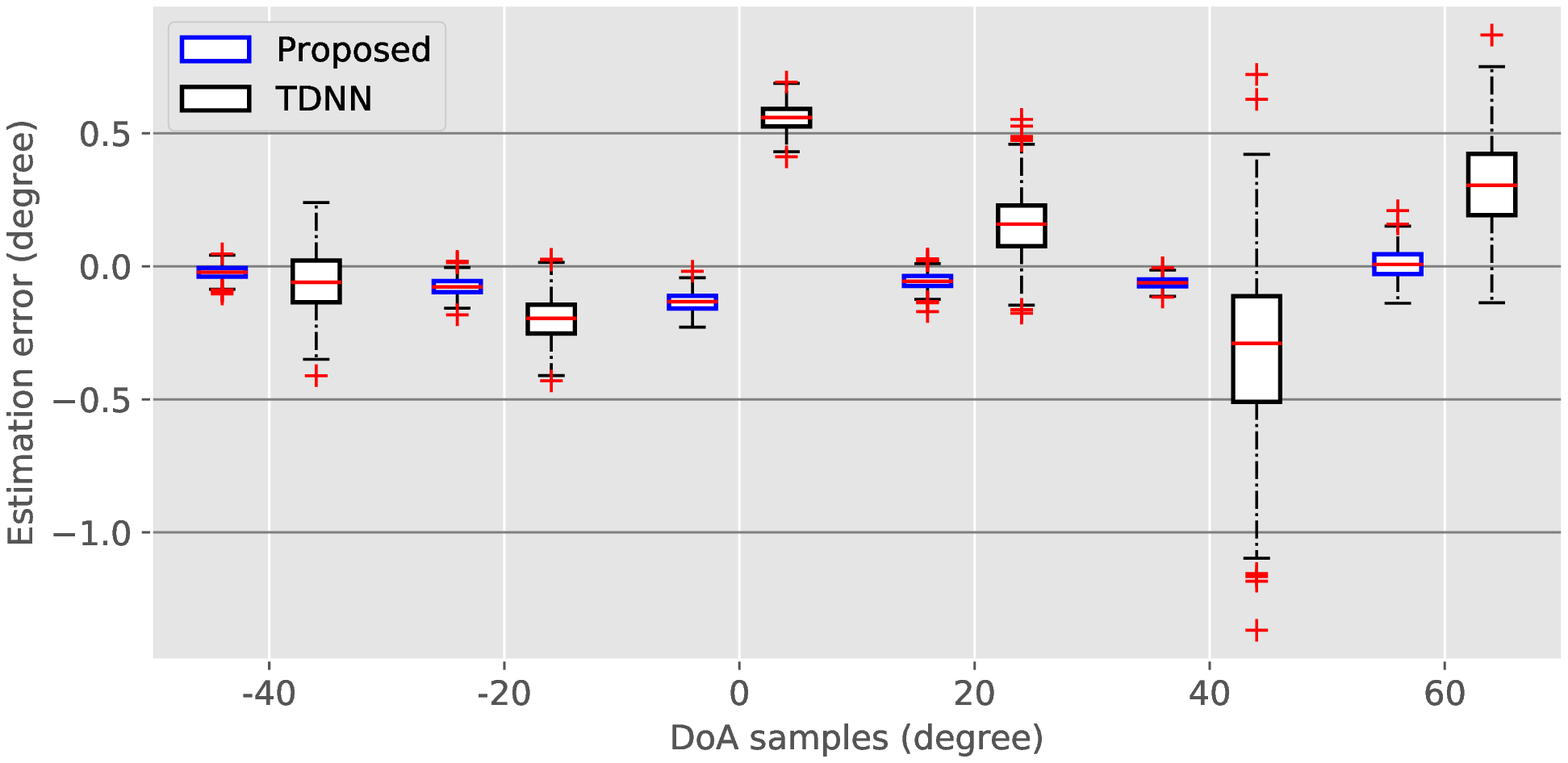}
	\caption{Box plot of degrees error for CVNN-based and TDNN-based DoA estimations.}
	\label{fig:box}
\end{figure*}

During the training process, the SGD method is employed to speed up the training, and the adaptive moment estimation (Adam) algorithm \cite{kingma2014adam} is used to update the parameters.
To investigate the loss functions that are appropriate for our regression predictive modeling problems, Fig. \ref{fig:loss} analyzes the effects of different loss metrics (i.e., MAE and MSE) on learning behavior and test performance.
Fig. \ref{fig:loss}(a) shows that both models converge to a steady state and the MSE loss are declining faster than the MAE loss. This is because the gradient of the MSE loss is high for larger loss values at initial epochs. The spikes appearing in the learning curve are the unavoidable consequences of the mini-batch GD methods.
We plot Fig. \ref{fig:loss}(b) to compare the test errors (in rad) trained with the MAE and MSE loss metrics. It is seen that the MAE-based test error can reach far below 0.05 after convergence, but the MSE-based test error is above 0.1. This validates the effectiveness and suitability of the MAE loss for our regression predictive task. 

In addition, we also study the impacts of residual blocks on the whole network performance. We train different networks with different numbers of residual blocks. Fig. \ref{fig:block} compares the MAE loss achieved by different networks. We see that the convergence rate improves with the number of residual blocks increasing from 1 to 3. The train error achieved by the 1-block network is worse than that of the other two networks. Although the 3-block network exhibits faster convergence than the 2-block network, the error performances of both two networks are comparable.

\subsection{Numerical Results}

In the following simulations, we mainly compare the performance of the proposed CVNN with that of the baseline methods under different conditions.

\begin{figure}[t]
    \centering
    \subfigure[]{\includegraphics[scale=0.4]{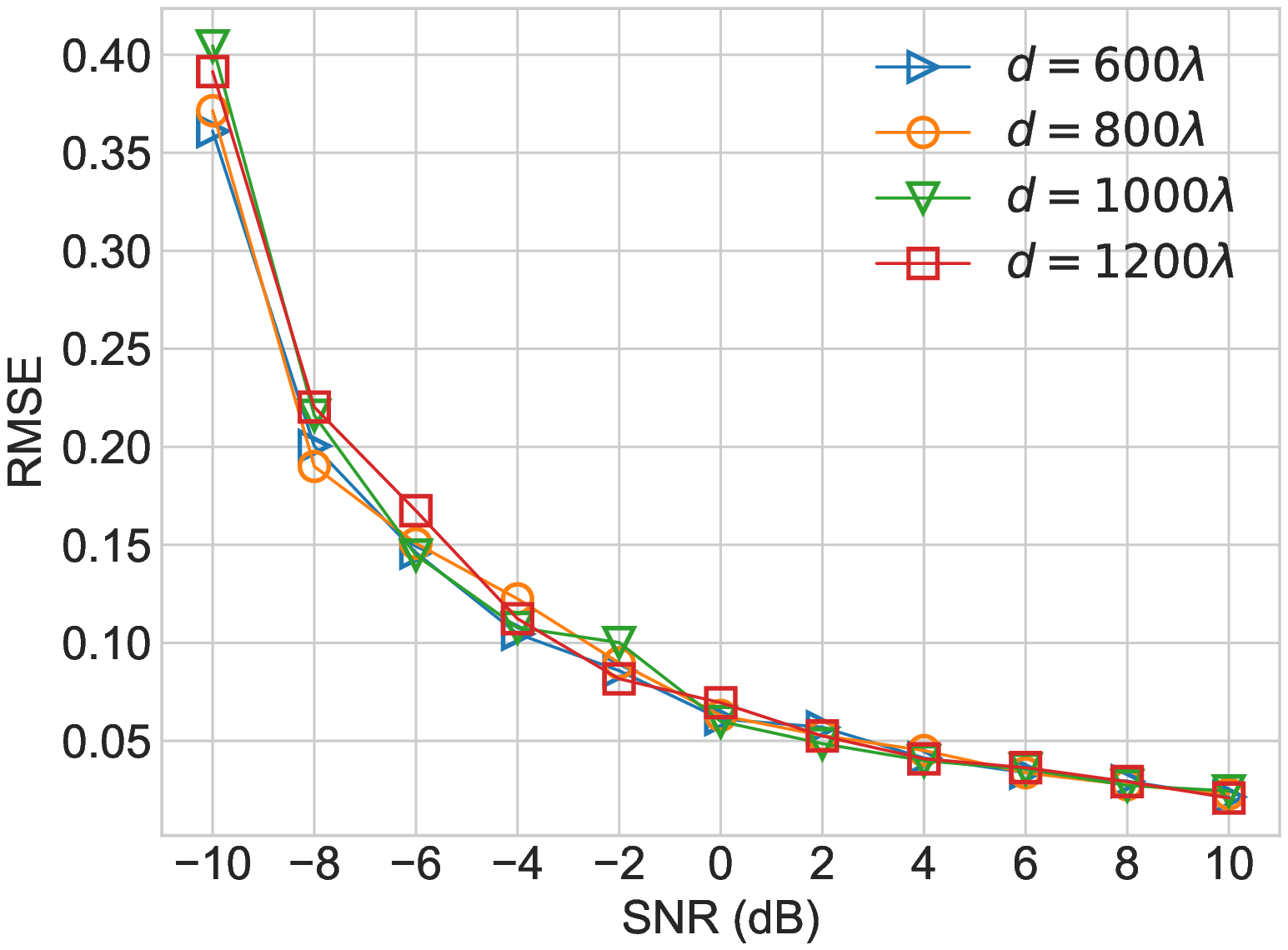}}
    \quad
    \subfigure[]{\includegraphics[scale=0.4]{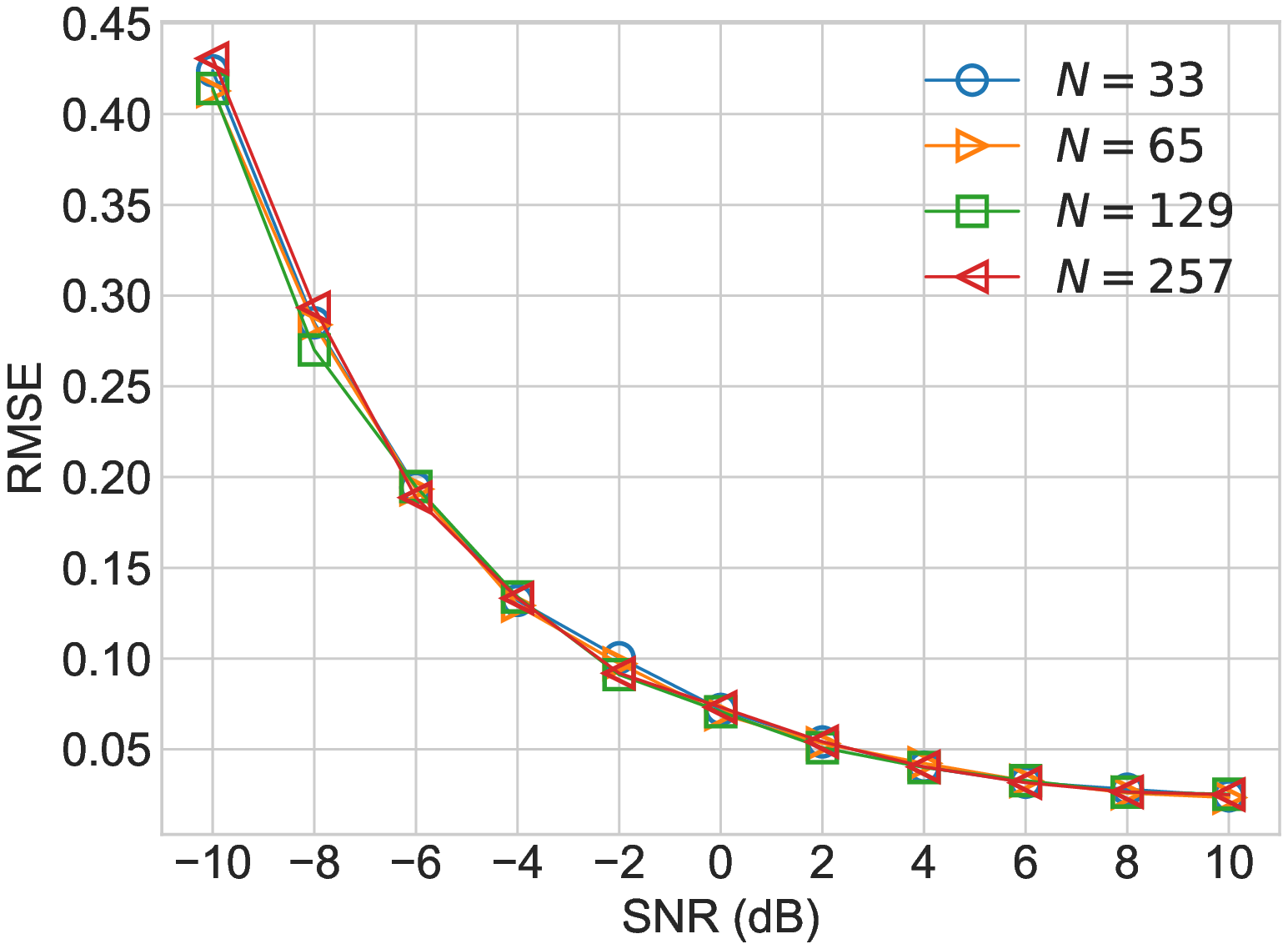}}
    \caption{The RMSE performance versus SNR for (a) different distances; (b) different numbers of antennas.}
    \label{fig:snr}
\end{figure}

First, a box plot is shown in Fig. \ref{fig:box} to explore the near-field DoA estimation performance in details, such as the distribution of predictive error.
We select 6 representative directions from $(-90^{\circ}, 90^{\circ})$, and obtain the corresponding 6 groups of test results by real and complex networks.
The SNR and the number of snapshots for all test samples are kept at 10 dB and 100, and these samples are placed on the circumference having a radius of $1000\lambda$ and centred on the reference array element.
In Fig. \ref{fig:box}, the boxes represent the interquartile range (IQR), and the red line on each box denotes the median. The whiskers (dashed lines outside the box) indicate the non-outlier range.
The outliers beyond the whiskers are shown as blue ``+''  symbols.
The box plots for both networks show the ranges and distributions of the degree errors obtained with 500 realizations.
We can see that the mean of degree errors achieved by the CVNN is closer to zero than that achieved by the TDNN in all directions.
Compared to the proposed CVNN method, there are significant large outliers for the TDNN estimates.
Besides, the IQR of the CVNN in each direction is narrower than that of the TDNN, which demonstrates that the CVNN is superior to the RVNN in terms of estimation robustness.
It should be noted that the RVNN does not take into account the correlation between the real and imaginary parts of the input features. In contrast, the internal structures of the complex-valued data are fully exploited by the proposed CVNN.

\begin{figure}[t]
    \centering
    \subfigure[]{\includegraphics[scale=0.4]{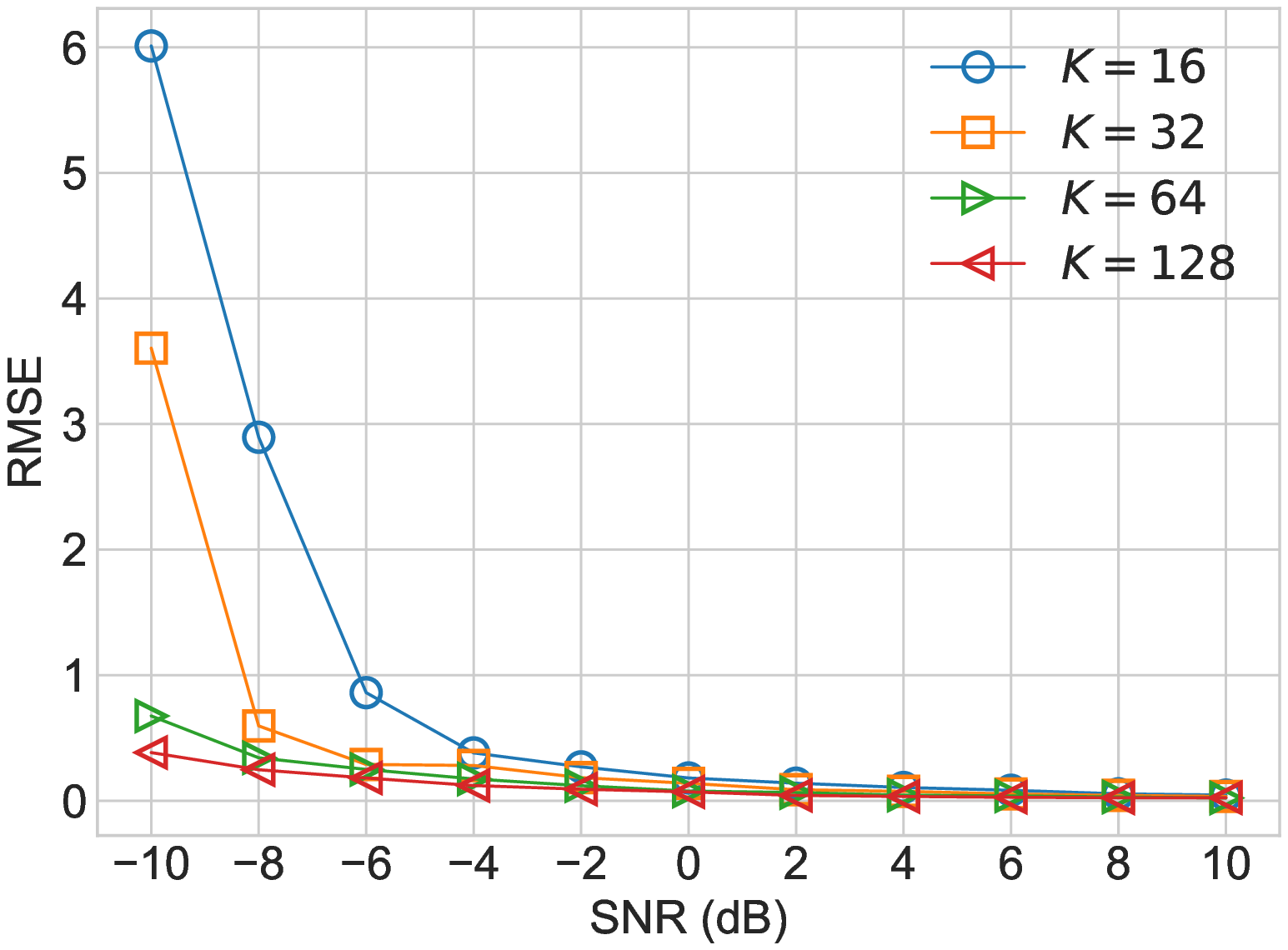}}
    \quad
    \subfigure[]{\includegraphics[scale=0.4]{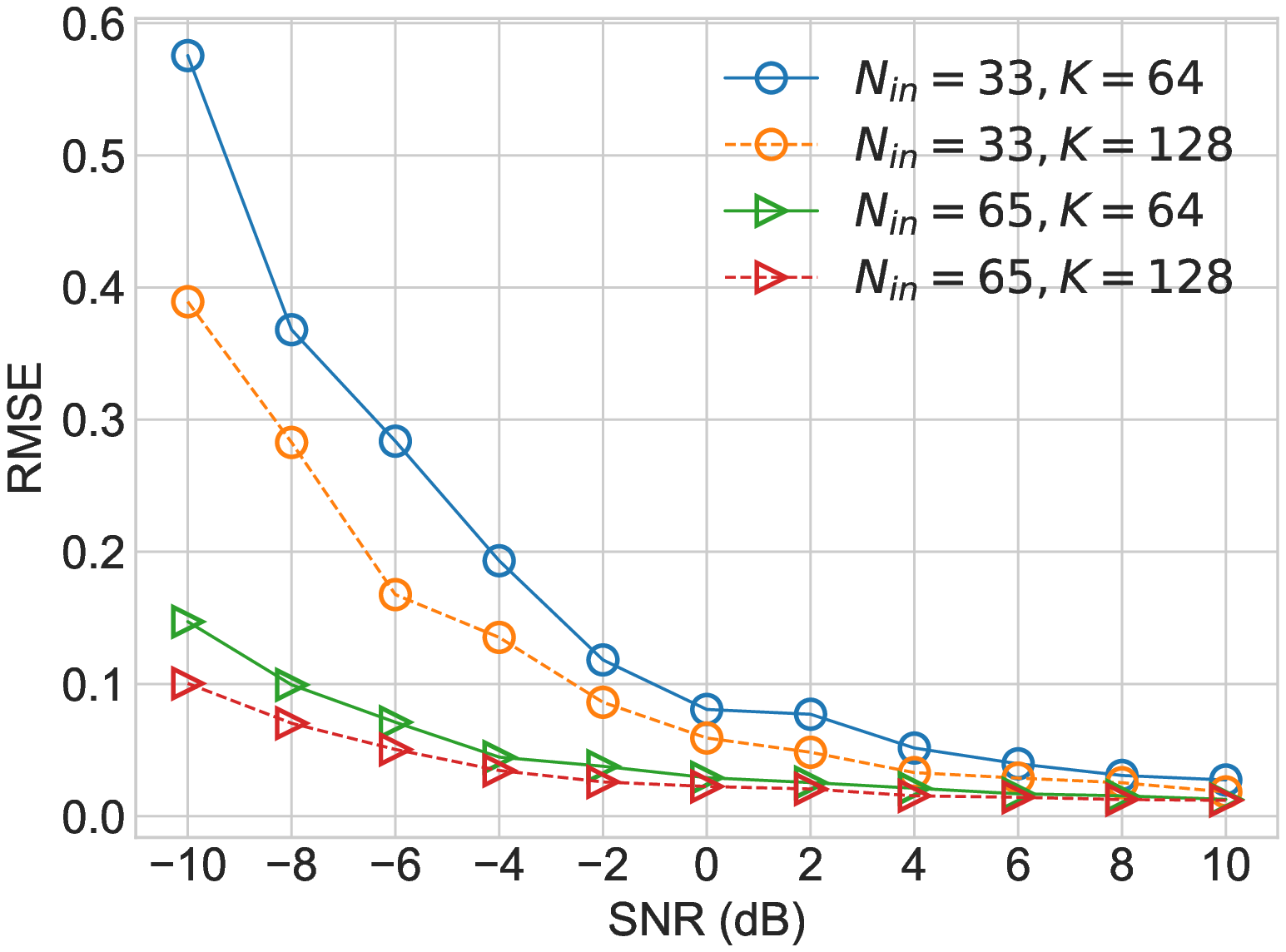}}
    \caption{RMSE performance versus SNR for (a) different numbers of snapshots; (b) different input sizes.}
    \label{fig:massive}
\end{figure}

\begin{figure}[t]
    \centering
    \subfigure[]{\includegraphics[scale=0.4]{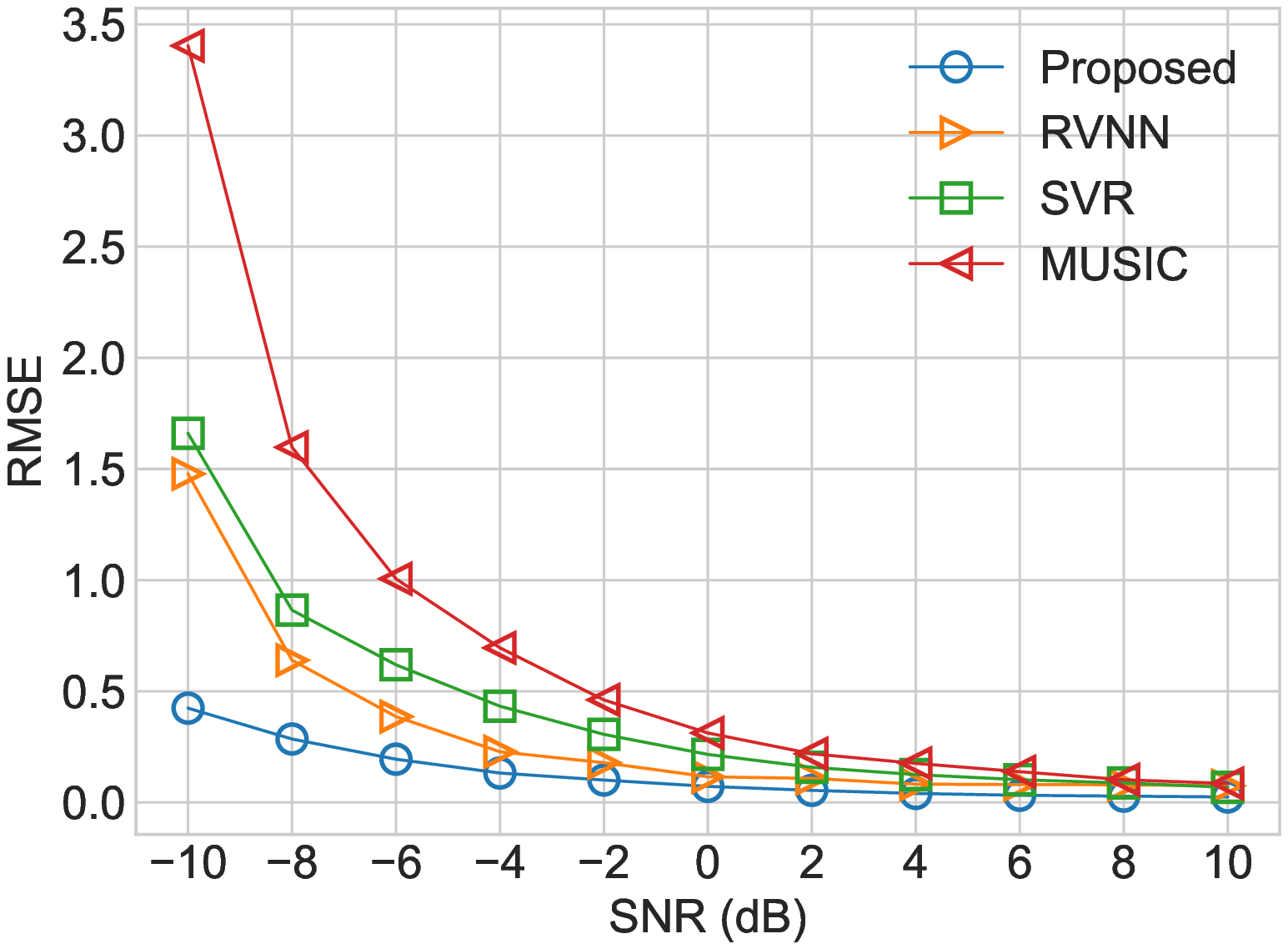}}
    \quad
    \subfigure[]{\includegraphics[scale=0.4]{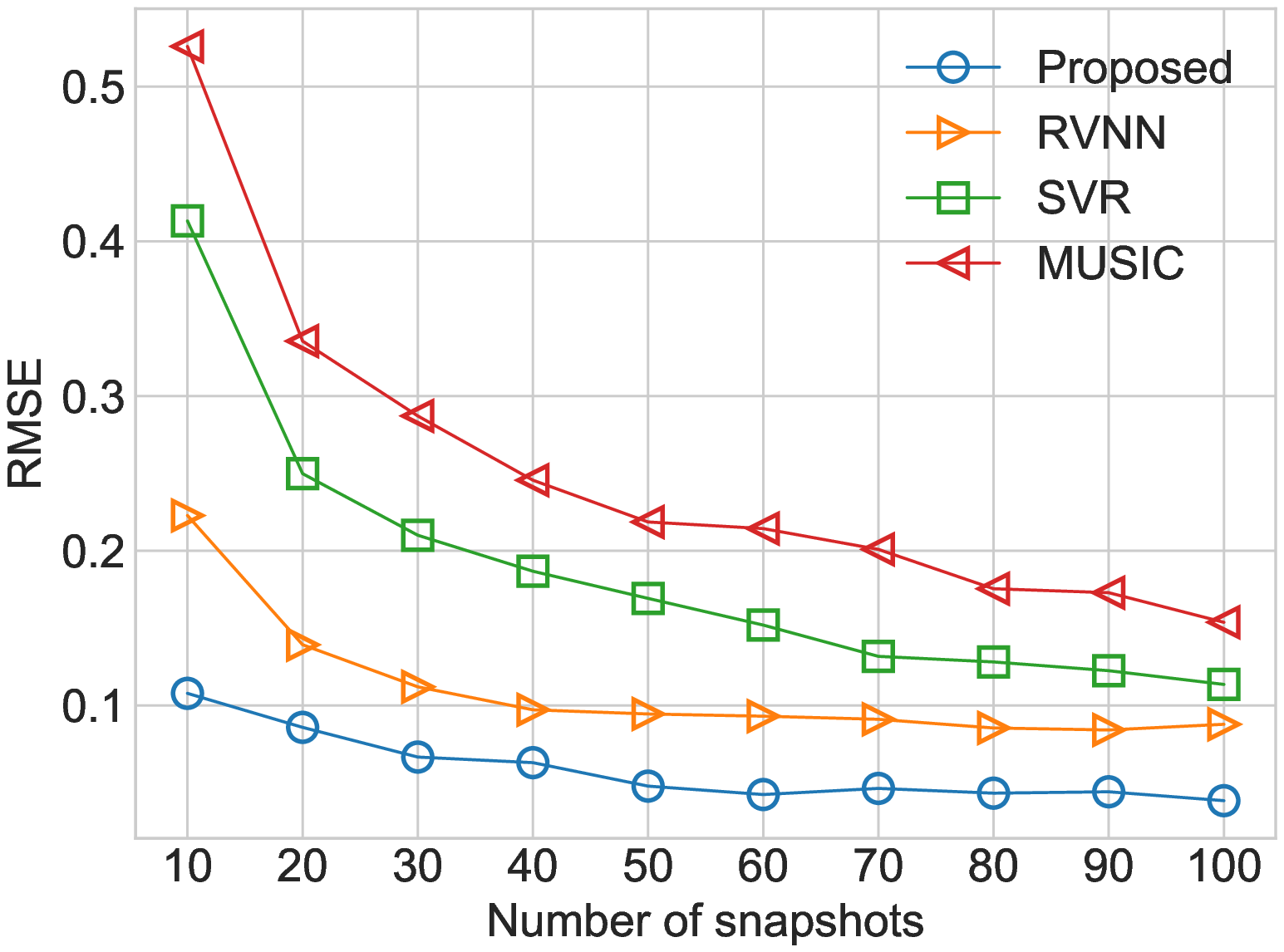}}
    \caption{RMSE performance for different methods versus (a) SNR; (b) number of snapshots.}
    \label{fig:comp}
\end{figure}

To test the generalization of the proposed model, we evaluate the RMSE performance under different distances and numbers of antennas.
Fig. \ref{fig:snr} shows the RMSE performance of the DoA estimation performed in 200 Monte Carlo simulations.
In Fig. \ref{fig:snr}(a), the signal sources are placed at different distances, including $600 \lambda$, $800 \lambda$, $1000 \lambda$, and $1200 \lambda$.
It is seen that, at different distances, there is almost no difference in the prediction performance of CVNN, which demonstrates that our proposed parameter decoupling method is effective.
The RMSEs of DoA decrease with the increasing SNR. Even at the worst SNR of -10 dB, it can still achieve the RMSE of about 0.4.
It is noted that the CVNN model offline trained at a fixed SNR of 10 dB can work well under other SNRs.
Next, Fig. \ref{fig:snr}(b) investigates the estimation performance of our scheme applied to different numbers of antennas. We observe that the RMSEs for all cases exhibit the same performance.
This indicates that the cropped VCM based scheme is feasible and flexible, since we do not need to redesign new networks for different input shapes with arbitrary antenna size.

Fig. \ref{fig:massive}(a) shows that the estimation performance tends to become better as the number of snapshots increases.
It is worth pointing out that the CVNN model offline trained with 100 snapshots can work well under the other number of snapshots.
We also notice that, with only 64 snapshots, the estimation performance can achieve the similar precision as that derived from more snapshots, which greatly reduces the latency in signal processing.
Further, Fig. \ref{fig:massive}(b) explores the performance of the CVNN scheme with respect to the input size $N_{\text{in}}$ based on the received data sampled from the MIMO system with 129 antennas.
We observe that, when the input size increases, the RMSE of our method decreases significantly. This is because the larger input size means that the received information of more antennas is utilized, thus avoiding the loss of original features. Moreover, the performance results in Fig. \ref{fig:massive} reveal that the DoA estimation precision is no longer sensitive to SNR when SNR is greater than -4 dB.

Fig. \ref{fig:comp} compares the performance between the proposed method and the baseline estimators.
In the first case shown in Fig. \ref{fig:comp}(a), we set $K=100$. Clearly, as the SNR varies from -10 to 10 dB, the RMSEs of all methods are gradually decreasing.
Note that the proposed CVNN outperforms the baseline schemes, because they call for relatively high SNRs to reach the low RMSE, while the RMSE of the proposed CVNN method can reach below 0.5 even at the worst SNR of -10 dB.
The RMSE comparison among different techniques at a fixed SNR of 5 dB is illustrated in Fig. \ref{fig:comp}(b). We see that the CVNN has the competitive performance when compared with the baseline methods with the same number of snapshots. Moreover, the CVNN method requires only 50 snapshots to reach the same level achieved by 100 snapshots. In conclusion, these results demonstrate the robustness of the proposed method to low SNR environments and the small number of snapshots.

\section{Conclusion}\label{section:con}
In this paper, a deep complex-valued network has been designed for near-field DoA estimation, which is modeled as an end-to-end regression problem. The proposed CVDL-aided DoA estimation framework can deal with complex-valued features directly. A VCM has been reconstructed to decouple the distance parameter from the angle parameter so that the angular variation information in the regression task can be learned smoothly. The 1-D convolutional ResNet with MAE loss has been designed for our specific regression task to enhance the training effects. Our simulation results have demonstrated that the CVDL model is superior to the baseline methods in terms of the performance of near-field direction finding. Furthermore, the proposed approach can work well when generalized to other SNRs and numbers of snapshots.

\bibliographystyle{IEEEtran}

\begin{IEEEbiography}[{\includegraphics[width=1in,height=1.25in,clip,keepaspectratio]{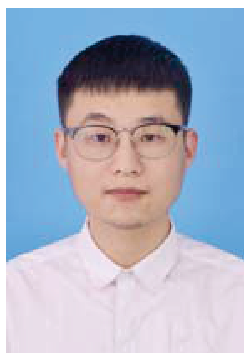}}]{Yashuai Cao}
(S'18) received the B.S. degree from Chongqing University of Posts and Telecommunications (CQUPT), Chongqing, China, in 2017. He is currently pursuing the Ph.D. degree in communication engineering with the School of Information and Communication Engineering, Beijing University of Posts and Telecommunications (BUPT), Beijing, China. His current research interests include communications and signal processing, massive MIMO and intelligent reflecting surface.
\end{IEEEbiography}
\begin{IEEEbiography}[{\includegraphics[width=1in,height=1.25in,clip,keepaspectratio]{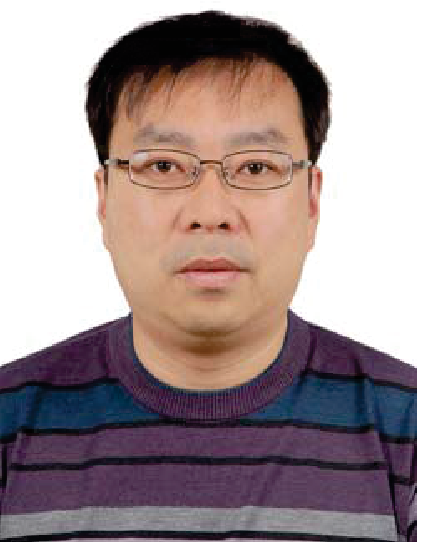}}]{Tiejun Lv}
(M'08-SM'12) received the M.S. and Ph.D. degrees in electronic engineering from the University of Electronic Science and Technology of China (UESTC), Chengdu, China, in 1997 and 2000, respectively. From January 2001 to January 2003, he was a Postdoctoral Fellow with Tsinghua University, Beijing, China. In 2005, he was promoted to a Full Professor with the School of Information and Communication Engineering, Beijing University of Posts and Telecommunications (BUPT). From September 2008 to March 2009, he was a Visiting Professor with the Department of Electrical Engineering, Stanford University, Stanford, CA, USA. He is the author of 2 books, more than 80 published IEEE journal papers and 180 conference papers on the physical layer of wireless mobile communications. His current research interests include signal processing, communications theory and networking. He was the recipient of the Program for New Century Excellent Talents in University Award from the Ministry of Education, China, in 2006. He received the Nature Science Award in the Ministry of Education of China for the hierarchical cooperative communication theory and technologies in 2015.
\end{IEEEbiography}
\begin{IEEEbiography}[{\includegraphics[width=1in,height=1.25in,clip,keepaspectratio]{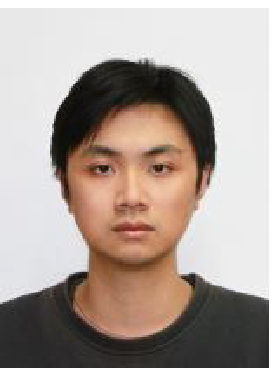}}]{Zhipeng Lin}
(S'17) is currently pursuing the dual Ph.D. degrees in communication and information engineering with the School of Information and Communication Engineering, Beijing University of Posts and Telecommunications, Beijing, China, and the School of Electrical and Data Engineering, University of Technology of Sydney, Sydney, NSW, Australia. His current research interests include millimeter-wave communication, massive MIMO, hybrid beamforming, wireless localization, and tensor processing.
\end{IEEEbiography}
\begin{IEEEbiography}[{\includegraphics[width=1in,height=1.25in,clip,keepaspectratio]{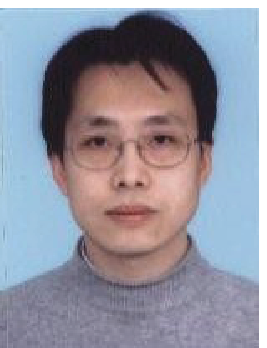}}]{Pingmu Huang}
received the M.S. degrees from Xi’an Jiaotong University, Xian, China, in 1996 and received Ph.D. degree of Signal and Information Processing from Beijing University of Posts and Telecommunications (BUPT), Beijing, China, in 2009. He is now a lecturer with the School of Information and Communication Engineering, BUPT. His current research interests include signal processing and pattern recognition. He published several journal papers and conference papers on signal processing and pattern recognition.
\end{IEEEbiography}
\begin{IEEEbiography}[{\includegraphics[width=1in,height=1.25in,clip,keepaspectratio]{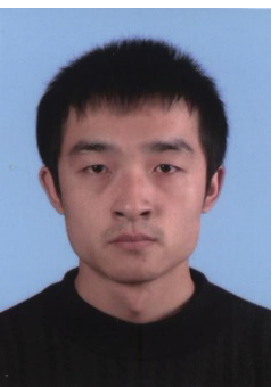}}]{Fuhong Lin}
received his M.S. degree and Ph.D. degree from Beijing Jiaotong University, Beijing, P. R. China, in 2006 and 2010, respectively, both in Electronics Engineering. Now he is a professor in department of Computer and Communication Engineering, University of Science and Technology Beijing, P. R. China. His research interests include Edge/Fog Computing, Network Security, and Big Data. He won ``Provincial and Ministry Science and Technology Progress Award 2'' in 2017 and 2019. His two papers won ``Top 100 most Cited Chinese Papers Published in International Journals'' in 2015 and 2016. He served as co-chair of the first and third IET International Conference on Cyberspace Technology, and general chair of the second IET International Conference on Cyberspace Technology. He was the leading editor of the Special issue ``Recent Advances in Cloud-Aware Mobile Fog Computing'' for Wireless Communications and Mobile Computing. Currently, he also serves as a reviewer more than 10 international journals including IEEE Transactions on Industrial Informatics, IEEE Access, Information Sciences, IEEE Internet of Things Journal, The Computer Journal and China Communications. He received the track Best Paper Award from IEEE/ACM ICCAD 2017.
\end{IEEEbiography}

\end{document}